\documentclass[fleqn,usenatbib]{mnras}

\usepackage{amsfonts}

\usepackage[T1]{fontenc}
\usepackage{ae,aecompl}
\usepackage{float}
\usepackage{comment}

\usepackage{grffile}
\usepackage{etoolbox}
\makeatletter 
  \patchcmd{\NAT@citex}
    {\@citea\NAT@hyper@{%
      \NAT@nmfmt{\NAT@nm}%
      \hyper@natlinkbreak{\NAT@aysep\NAT@spacechar}{\@citeb\@extra@b@citeb}%
      \NAT@date}}
    {\@citea\NAT@nmfmt{\NAT@nm}%
    \NAT@aysep\NAT@spacechar\NAT@hyper@{\NAT@date}}{}{}

  \patchcmd{\NAT@citex}
    {\@citea\NAT@hyper@{%
      \NAT@nmfmt{\NAT@nm}%
      \hyper@natlinkbreak{\NAT@spacechar\NAT@@open\if*#1*\else#1\NAT@spacechar\fi}%
        {\@citeb\@extra@b@citeb}%
      \NAT@date}}
    {\@citea\NAT@nmfmt{\NAT@nm}%
    \NAT@spacechar\NAT@@open\if*#1*\else#1\NAT@spacechar\fi\NAT@hyper@{\NAT@date}}
    {}{}
\makeatother

\usepackage{etoolbox}
 \makeatother
\usepackage{graphicx}	
\usepackage{amsmath}	
\usepackage{amssymb}	
\usepackage{pifont}
\newcommand\HII{{H\,\textsc{ii}}} 
\newcommand\CII{{C\,\textsc{ii}}}
\newcommand\OI{{O\,\textsc{i}}}
\title[Stellar wind enrichment from Pop~III spinstars]{Stellar winds and metal enrichment from fast-rotating Population~III stars}

\author[B. Liu et al.]{Boyuan Liu\textsuperscript{\href{https://orcid.org/0000-0002-4966-7450}{\includegraphics[width=2.5mm]{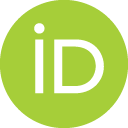}}\,}\thanks{E-mail: boyuan@utexas.edu}$^{1}$, 
Yves Sibony$^{2}$,
Georges Meynet$^{2}$
and Volker Bromm$^{1}$
\\
$^{1}$Department of Astronomy, University of Texas, Austin, TX 78712, USA\\
$^{2}$Geneva Observatory, University of Geneva, Chemin des Maillettes 51, 1290 Sauverny, Switzerland
}

\date{Accepted XXX. Received YYY; in original form ZZZ}

\pubyear{2020}

\begin{document}
\label{firstpage}
\pagerange{\pageref{firstpage}--\pageref{lastpage}}
\maketitle

\begin{abstract}
Stellar winds from fast-rotating Population~III (Pop~III) stars have long been suspected to make important contributions to early metal enrichment, as features in the nucleosynthesis of such `spinstars' are consistent with the chemical abundance patterns of some metal-poor stars in the local Universe. Particularly, stellar winds rich in light elements can provide another pathway towards explaining the carbon enhancement in carbon-enhanced metal-poor (CEMP) stars. In this work, we focus on the feedback of Pop~III stellar winds combined with supernovae (SNe), and derive the resulting chemical signatures in the enriched medium. We explore a large parameter space of {Pop~III star formation, feedback, yields from winds and SNe} with a semi-analytical model. 
The predicted pattern of carbon and iron abundances of second-generation stars agrees well with observations of CEMP-no stars ($[\rm Ba/Fe]<0$) at $[\rm Fe/H]\lesssim -3$ and $A(\mathrm{C})\lesssim 7$, under the optimistic assumption of significant mass loss by winds {from massive ($\gtrsim 25\ \rm M_{\odot}$) stars that collapse into BHs without SNe}. In this scenario, carbon-rich but iron-free second-generation stars can form in systems dominated by enrichment from winds, gaining trace amounts of iron by accretion from the interstellar medium, to become the most iron-poor and carbon-enhanced stars seen in observations ($[\rm Fe/H]\lesssim -4$, $[\rm C/Fe]\gtrsim 2$). {We conclude that the observed CEMP-no stars can be explained by both our winds + ISM accretion channel as well as the well-studied faint SN scenario.} Wind feedback from Pop~III spinstars deserves more detailed modelling in early cosmic structure formation. 
\end{abstract}



\section{Introduction}
\label{s1}
The first generation of stars, the so-called Population III (Pop III), formed in primordial gas with extremely low or zero metallicity ($Z\lesssim 10^{-6}-10^{-4}\ \rm Z_{\odot}$), have distinct features compared with present-day, Population~I (Pop~I), stars (see e.g. \citealt{bromm2013,haemmerle2020formation} for reviews). On the one hand, they follow very different stellar evolution tracks, being more compact and hotter than stars with solar composition. On the other hand, due to insufficient cooling of primordial gas, they are more massive than present-day stars, with a top-heavy initial mass function (IMF). In the absence of direct observations, which is likely still challenging in the \textit{James Webb Space Telescope (JWST)} era (e.g. \citealt{gardner2006james,anna2020tele}), detailed properties of Pop~III stars remain uncertain from theoretical predictions, such as the exact form of their IMF (e.g. \citealt{susa2014mass,hirano2015primordial,machida2015accretion,stacy2016building,hirano2017formation}), binary properties (e.g. \citealt{stacy2013constraining,hosokawa2020,liu2021dynamical}), and their fates/remnants (e.g. \citealt{heger2010nucleosynthesis,tanikawa2020fitting}). Nevertheless, constraints have been inferred from indirect observations, such as stellar archaeology {(reviewed by \citealt{frebel2015near}) for extremely metal-poor (EMP) stars in the Milky Way (MW) stellar halo and dwarf satellites (e.g. \citealt{tumlinson2006,salvadori2007,hartwig2015,salvadori2015,debennassuti2017,magg2019observational,ishigaki2018initial,salvadori2019,yuta2020,rossi2021})}, 21-cm cosmology (e.g. \citealt{schauer2019constraining,qin2020,chatterjee2020}), and gravitational waves from mergers of compact objects (e.g. \citealt{kinugawa2014possible,hartwig2016,belczynski2017likelihood,tanikawa2020merger,boyuan2020,bl2020gw190521}).

Among those indirect probes, stellar archaeology has so far provided the most robust constraints on the Pop~III IMF and supernova (SN) properties from the abundance patterns in metal-poor stars. A peculiar stellar population called carbon-enhanced metal-poor (CEMP) stars (with $\rm [C/Fe]>0.7$ and $\rm [Fe/H]\le-1$; see e.g. \citealt{beers2005discovery,aoki2007carbon}) has received particular attention. Their fraction is higher at lower [Fe/H] {(e.g. \citealt{frebel2010,norris2013most,spite2013,carollo2014carbon,bonifacio2015})}, implying that carbon enhancement is a feature of metal enrichment from the first stars. Particularly, a subgroup of CEMP stars without enhancement of neutron-capture elements, the so-called CEMP-no stars ($[\rm Ba/Fe]<0$)\footnote{In difference, CEMP-$s$ stars are widely believed to originate from mass transfer from intermediate mass companion stars, as almost all CEMP-$s$ stars are in binaries (e.g. \citealt{lucatello2005binary,masseron2010,abate2013wind,starkenburg2014binarity}).}, are expected to be the bona-fide second-generation (Pop~II) stars that preserve the chemical signatures of Pop~III metal-enrichment {(for details, see e.g. \citealt{yoon2016observational,hansen2019abundances,dietz2020two})}. 

The detailed mechanisms of forming CEMP-no stars are still under debate. Several scenarios are proposed, such as `faint SNe' or jet-induced SNe with fallback and mixing {(e.g. \citealt{maeda2003bipolar,umeda2003first,umeda2005variations,iwamoto2005,tominaga2008aspherical})}, where most iron-peak elements in the inner region fall back onto the central compact object, while the carbon-rich outer layer is ejected. Other models invoke binary mass transfer (e.g. \citealt{suda2004he,komiya2007origin}), where additional carbon is gained from a companion star similar to the case of CEMP-$s$ stars, and fast-rotating massive Pop~III stars, the so-called `spinstars' (e.g. \citealt{meynet2006early,hirschi2007very,ekstrom2008effects,yoon2012evolution,choplin2017some,choplin2020nucleosynthesis}), whose strong winds can pollute the surroundings with carbon (and other light elements). 

The first two scenarios have been extensively studied in the context of cosmic structure formation with semi-analytical models and cosmological simulations {(e.g. \citealt{cooke2014carbon,salvadori2015,sarmento2016following,debennassuti2017,sharma2018origins,hartwig2019fingerprint,komiya2020faint,chiaki2020seeding,jeon2021role})}, which can indeed reproduce the observed distribution of CEMP-no stars in the carbon enhancement and iron abundance space (with certain parameter choices). On the contrary, the metal enrichment process driven by Pop~III stellar winds has not been investigated in any detail yet. More complete analysis is required, as it has recently been shown by \citet{magg2020minimum} that the dilution mass into which metals are mixed is crucial for inferring the properties of Pop~III from stellar archaeology. 

Actually, it is suggested that spinstars not only produce carbon enhancement, but also exhibit other interesting chemical signatures \citep{chiappini2013first,maeder2015first},
such as the production of primary $^{14}$N \citep{MM2002,Chiappini2003, hirschi2007very}, $^{13}$C \citep{Chiappini2008}, $^{22}$Ne, and $s$-process elements \citep{Pigna2008, chiappini2011imprints, Cescutti2013, Frisch2016, Sique2016, Choplin2018,Limongi2018}. Spinstars also impact the evolution of the C/O ratio at very low metallicities \citep{Chiappini2006} and may allow the production
of primary Be and B by spallation reaction in their winds \citep{Prantzos2012}. Some stars produced by such spinstars may also be He-rich \citep{GM2010}. { At zero metallicity, not all these outputs are present. For instance, it has been shown by \citet{Frisch2016} that
production of $s$-process elements in spinstars only occurs for metallicities larger than about 10$^{-3}$ Z$_\odot$. Below that metallicity the abundance of iron seed nuclei is too low for allowing a significant production of neutron-rich isotopes.}

Moreover, it is found in hydrodynamic simulations \citep{stacy2011rotation,stacy2013rotation} that Pop~III protostars rotate at very high rates (50-100\% of the critical/Keplerian rate), so that fast-rotating stars are likely common in Pop III, especially considering the weakness of magnetic fields in the early Universe \citep{hirano2018angular}. The stellar winds from such stars can be important for metal enrichment, particularly for massive stars with $m_{\star}\sim 25-120\ \rm M_{\odot}$, which are expected to collapse completely into black holes (BHs) without SN explosions. 

In the light of this, we study the feedback of stellar winds from fast-rotating Pop~III stars in combination with SN feedback, focusing particularly on the carbon and iron abundances in the enriched medium that hosts second-generation stars. We couple the Pop~III stellar evolution grids from \citet{murphy2021grids} with an idealized semi-analytical model of metal enrichment from Pop~III stellar winds and SNe in self-enriched minihaloes and atomic-cooling (AC) haloes, the typical sites of Pop~III star formation. A large parameter space for Pop~III star formation and feedback is explored to evaluate whether the spinstar scenario can reproduce the observed CEMP-no stars in terms of carbon and iron abundances, and how the signatures of Pop~III metal enrichment depend on input parameters (e.g. wind properties and IMF). 

This paper is organised as follows. In Section~\ref{s2}, we discuss the interplay between rotation and winds for Pop III stars and introduce our stellar evolution grids of non-rotating and rotating Pop~III stars, based on which metal yields from winds and SNe are extracted (Sec.~\ref{s2.1}). 
We also describe our metal enrichment model for both wind and SN feedback (Sec.~\ref{s2.2}), as well as the construction of Pop~III clusters (as basic units of Pop~III star formation) and derivation of the relevant feedback parameters that are fed into the metal enrichment model (Sec.~\ref{s2.3}). In Section~\ref{s3}, we then present the chemical signatures of second-generation stars formed in self-enriched minihaloes (Sec.~\ref{s3.1}) and AC haloes (Sec.~\ref{s3.2}). Finally, Section~\ref{s4} summarizes our main findings and implications for future studies.

\section{Methodology}
\label{s2}
In this section, we describe our approach of calculating the chemical patterns of the interstellar medium (ISM) enriched by stellar winds and SNe from Pop~III stars, which are inherited by the second generation of stars, potentially observed as EMP stars in the local Universe. We start with a brief discussion on Pop~III stellar evolution and winds, introducing our grids of stellar evolution models for Pop~III stars with and without rotation \citep{murphy2021grids}, as well as the relevant metal yields from winds and SNe (Sec.~\ref{s2.1}). We then build a simple semi-analytical enrichment model to estimate the dilution mass for the produced metals (Sec.~\ref{s2.2}), given the input parameters for stellar feedback and the final host halo of the second generation of stars. Finally, we describe our model for Pop~III star formation (Sec.~\ref{s2.3}), which serves as the bridge between stellar evolution grids and feedback parameters for metal enrichment. Throughout our calculations, we adopt the \textit{Planck} cosmological parameters for $\Lambda$CDM, $\Omega_{\rm m}=0.3089$, $\Omega_{\rm b}=0.048$ and $h=0.6774$ \citep{planck}, and solar overall, carbon and iron mass fractions, $\rm Z_{\odot}=0.02$, $Y_{\rm C,\odot}\simeq 3.3\times 10^{-3}$ and $Y_{\rm Fe,\odot}\simeq 1.7\times 10^{-3}$. 

\subsection{Pop~III stellar evolution}
\label{s2.1}


\begin{figure*}
    \centering
    \includegraphics[scale=.49]{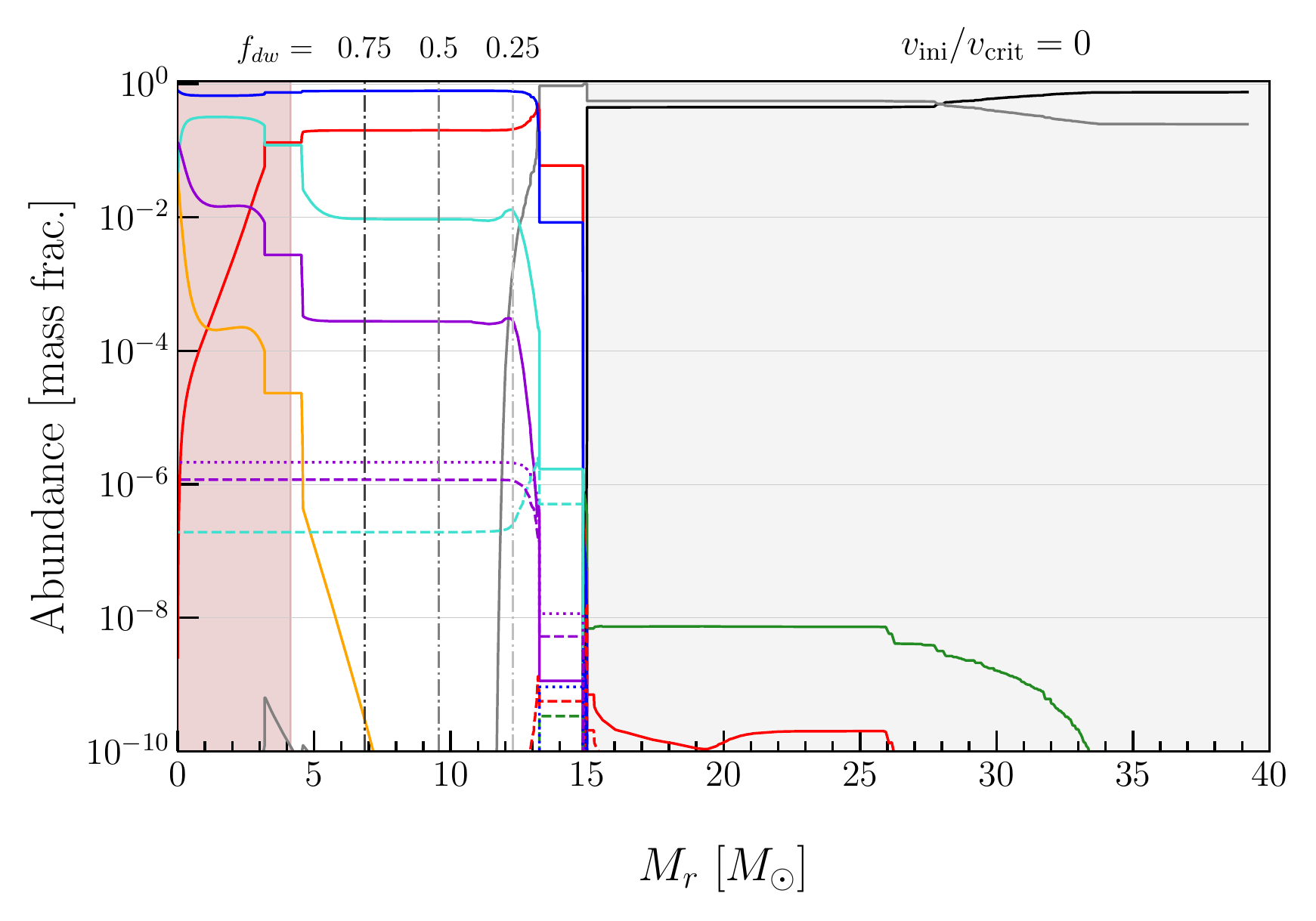}    \includegraphics[scale=.49]{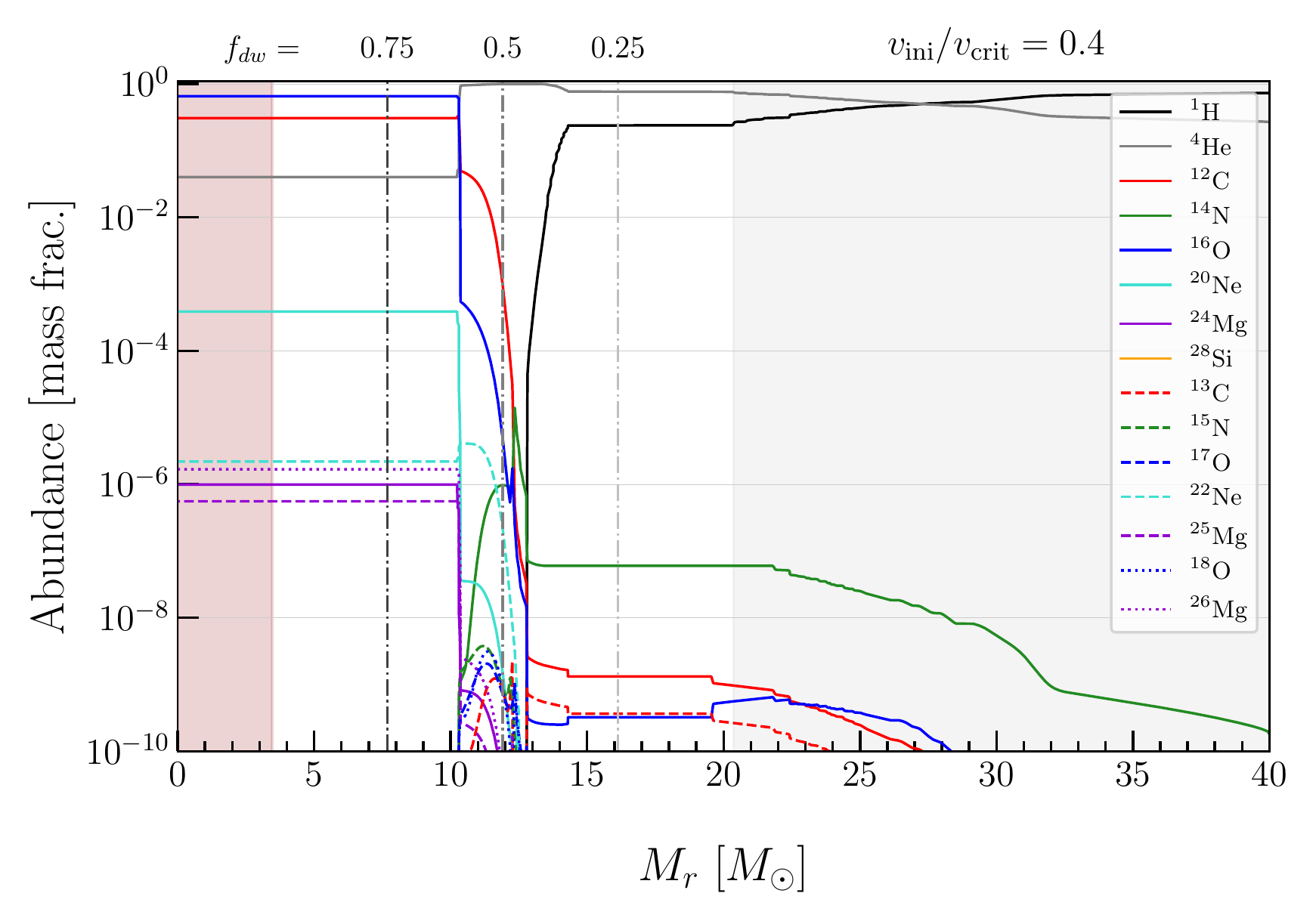}
    \vspace{-10pt}
    \caption{Abundances (mass fractions) of selected elements as a function of the Lagrangian mass coordinate at the end of the Ne-photodisintegration phase and close to the end of He-burning (at a core He mass fraction $Y_{\rm c}=0.04$) for the 40~M$_\odot$ Pop~III non-rotating (left) and rotating (right) models, respectively. 
    The solid black line is for $^1$H, the grey solid line for
    $^4$He, the solid (dashed) blue, red and green lines are for $^{16}$O ($^{17}$O), $^{12}$C ($^{13}$C), and $^{14}$N ($^{15}$N) respectively. The cyan and magenta solid (dashed) lines are for $^{20}$Ne ($^{22}$Ne), and $^{24}$Mg ($^{25}$Mg), respectively. The dotted lines (blue and magenta) are for $^{18}$O and $^{26}$Mg. Finally the orange solid lines are for $^{28}$Si. The pink shaded area corresponds to the stellar remnant (left after SN) based on the CO core-remnant mass relation in \citet{Limongi2012}. The grey shaded area corresponds to the minimum mass ejected by winds. The dashed vertical lines in between the pink and grey shaded areas, labeled by different values of $f_{dw}$, show the corresponding masses left after wind mass loss.}
    \label{fig:chemie}
\end{figure*}

The evolution of Pop III stars differs in many ways from that of stars containing even a very small amount of metals \citep[see e.g.][and references therein]{Heger2001,Marigo2003, Ekstrom2006,Limongi2012, Yoon2012, murphy2021grids}. However there is a common feature
of all stars with no or only traces of heavy elements: they do not lose any significant  mass by line-driven stellar winds \citep{Krti2008}. 
This is expected since metal-poor material presents only very few absorption lines and thus is not efficient in transforming part of the radiative energy into bulk kinetic energy \citep{Kud2002}.
However, this conclusion can be different in rotating stellar models
as we shall see below \citep{meynet2006early,hirschi2007very,ekstrom2008effects},
and/or if mass loss is triggered through processes that are not or weakly dependant on metallicity, such as some kinds of pulsations or instabilities triggered by internal gravity waves \citep{smith2006role,van2008numerical, Yoon2010,Fuller2017, Fuller2018,Fuller2020, Fuller2021}.

Stellar rotation changes the mass lost by winds in many different ways. First it increases the line-driven winds for a star at a given position in the Hertzsprung-Russell (HR) diagram \citep{maeder2000stellar}. This effect,
however, is significant only if the non-rotating star would have already significant line-driven stellar winds, which is not the case for Pop III stars. More important for very metal-poor stars is the fact that rotation modifies the chemical structure of a star, allowing some heavy elements produced in the core by nuclear burning to reach the surface. Thus even if the star had begun its evolution with no metals in the outer layers, during its evolution, the surface might show significant abundances of  heavy elements due to this rotationally-induced self-enrichment process. A particularly
extreme situation would be the case of stars following a homogeneous evolution due to a very efficient internal mixing \citep{Maeder1987, YL2005, Song2016}.

These surface enrichments may then trigger line-driven winds.
This effect has been studied in the case of very metal-poor massive stars and may provide an explanation for the origin of CEMP-no stars \citep{meynet2006early,hirschi2007very}. The mass lost can be significant. As a numerical example, \citet{meynet2006early} find
that a 60 M$_\odot$ star at $Z$=10$^{-8}$, with an initial rotation of
800 km s$^{-1}$, can lose nearly 2/3 of its initial mass through this process, leaving a final mass of only 24 M$_\odot$. Similarly, \citet{hirschi2007very} argues, using the same mechanism, that a 85 M$_\odot$ star with the same initial metallicity and rotation as the 60 M$_\odot$ case mentioned above will lose 65 M$_\odot$, more than 75\% of its initial mass. As a further example, \citet{EkFIRST2008} show that a fast rotating Pop~III 150 M$_\odot$ model, taking into account the Taylor-Spruit dynamo, loses 60\% (90 M$_\odot$) of its
initial mass. Interestingly, since the mechanism that triggers the winds in these models is due to the presence of newly synthesized elements in the star, this material has also a direct impact on the chemical enrichment of the surrounding interstellar medium. As mentioned above, this scenario has been suggested as one pathway for the origin of CEMP-no stars.
Unfortunately, at the moment, stellar wind models with the peculiar abundances created by rotation, and for the temperatures and luminosities where this process occurs, are still missing. In the above models, the mass loss has been simply increased by a factor $(Z_{\rm surf}/Z_\odot)^{0.5}$, where
$Z_{\rm surf}$ is the actual surface metallicity, using the same
metal dependence as for the initial metallicity as suggested by \citet{Kud1987}.


Rotation has also an impact on the winds triggered by the continuum rather than by lines. Indeed, by adding a centrifugal acceleration, rotation changes the expression of the Eddington limit \citep{maeder2000stellar}, imposing a lower maximum luminosity of a star. Above that limit, the continuum radiation is able to launch stellar winds. This may be a significant effect for luminous stars. 
Even far from the Eddington limit, rotation may induce a mechanical wind when at the surface, the centrifugal acceleration balances the gravity at the equator \citep{ekstrom2008effects}. This mechanical mass loss is likely much more frequent at very low metallicity or in Pop III stars, since in those stars, the angular momentum brought to the surface by internal angular momentum processes, is not removed by line-driven winds.

Regardless of rotation, some other effects may be responsible for mass loss episodes during the evolution of massive stars, rather independent of metallicity, and thus occuring even in Pop~III or very metal poor stars. An example may be the mechanism responsible for the strong mass loss experienced by Luminous Blue Variables
\citep{smith2006role, Zhao2020, Grassi2021}.
Pulsation, either induced when the star is in a red supergiant stage \citep{Yoon2010}, or in a blue supergiant phase \citep{Saio2013}, may be linked to violent mass loss episodes. The most massive stars may undergo pulsational pair instabilities \citep{Woosley2007}.
Some authors have suggested that the action of gravity waves at the very end of the evolution of massive stars may trigger substantial mass loss \citep{Fuller2017,Fuller2018,Fuller2020,Fuller2021}, providing a possible explanation for why so many observed supernovae show signs of an interaction between the supernova ejecta and a wind launched just before the collapse \citep{Strot2021}. 

For most of the above processes, a detailed description of the physics responsible for these events is still missing, and only a few exploratory works have been performed so far (see references above). In the present study, we wish to explore
whether the consequences for the chemical enrichment of the ISM by
the ejection of material from Pop~III stars via winds are significantly different
from those due to supernova explosions. In view of the large uncertainties concerning the importance of winds, we adopt a parametric approach where the wind intensity can be varied between some limits (see below). While such an approach will not provide a secure modeling of the winds, it can at least provide an indication whether winds have the potential of modifying the classical picture of no-winds for Pop III stars. If yes, we hope that this work will trigger subsequent works where the physical basis for these winds will be thoroughly investigated.

In the present work, we used the recent Pop III stellar models by \citet{murphy2021grids}. {These models contain non-rotating ($v_{\rm ini}=0$) and rotating cases ($v_{\rm ini}/v_{\rm crit}=0.4$) where $v_{\rm ini}$ is the initial equatorial rotation velocity, and $v_{\rm crit}$ is the break-up velocity at critical rotation (see equs.~3 and 4 in \citealt{hirschi2007very}).} Only the effects of line-driven winds and of mechanical mass loss have been accounted for in these models. Both these effects have been found to remain very modest. However, as explained above, we may miss important mass loss processes such as winds triggered by surface self-enrichment in response to more effective rotational mixing than accounted for here, by continuum radiation, by pulsations, and/or by gravity wave instabilities. To explore the potential impact of wind mass loss, we need to assign values to the following quantities: the mass that is lost, the timescale of the winds, as well as the velocities and chemical compositions. In the following, we describe our procedures to do so.

\begin{figure}
    \centering
    \includegraphics[scale=.45]{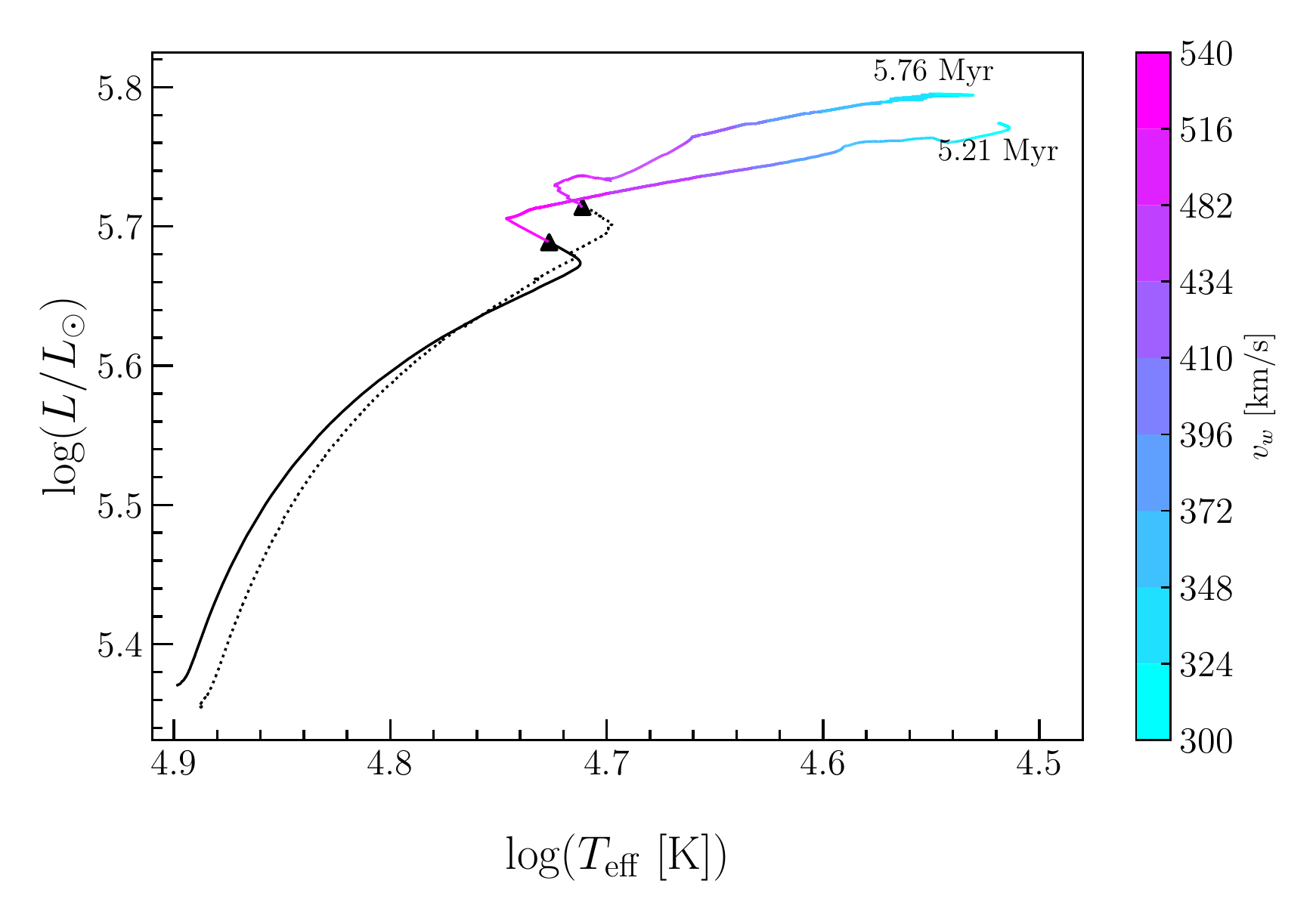}
    \vspace{-10pt}
    \caption{Evolutionary tracks of the rotating (dotted) and non-rotating (solid) 40 M$_\odot$ models in the HR diagram. The end points of the Main Sequence (triangles) are indicated, as well the ages at the end of evolution. The tracks are color coded with the wind velocities given by Equ.~\ref{e14} for the post-MS phase.}
    \label{fig:HRD_P020z00}
\end{figure}

\begin{figure*}
    \centering
    \includegraphics[scale=.49]{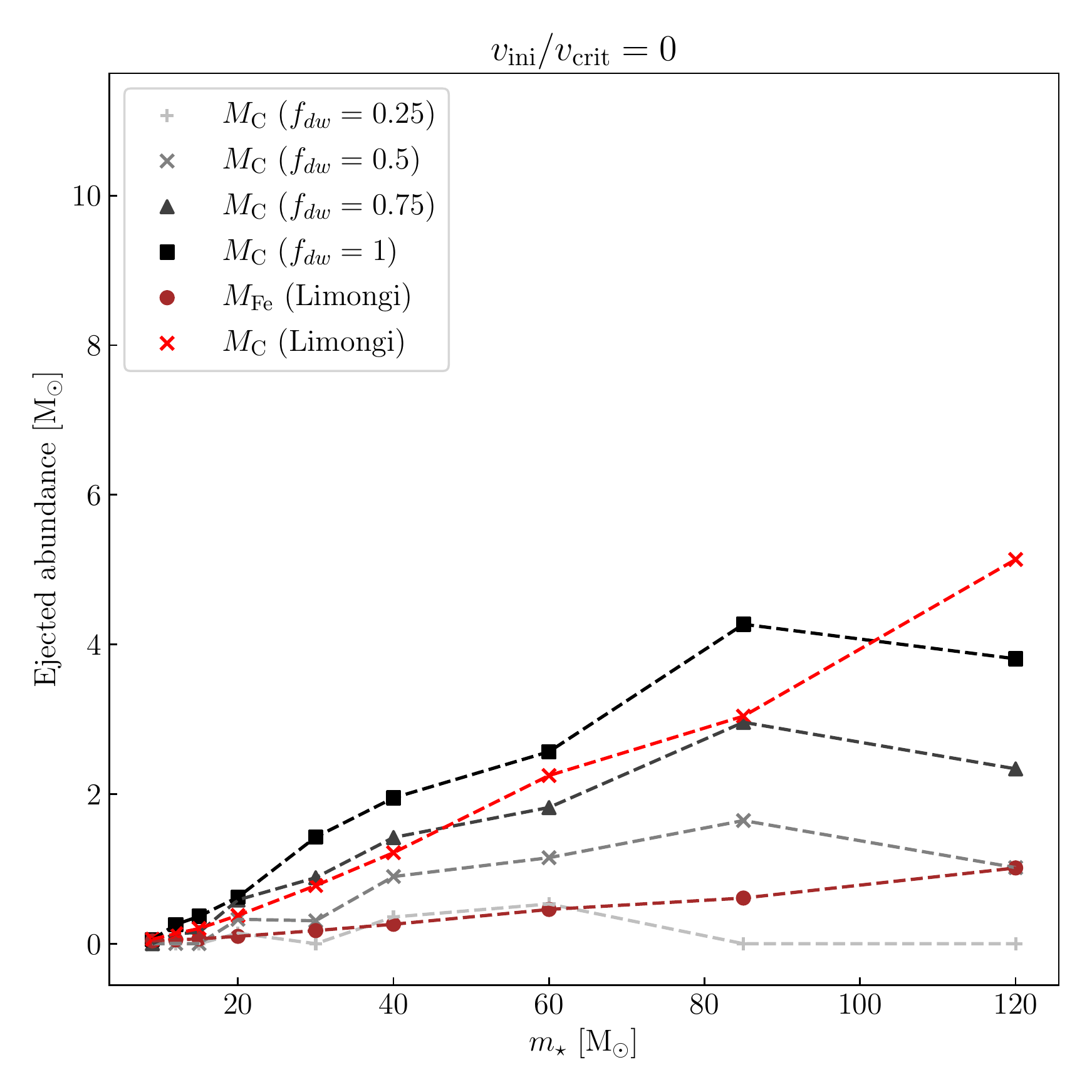}    \includegraphics[scale=.49]{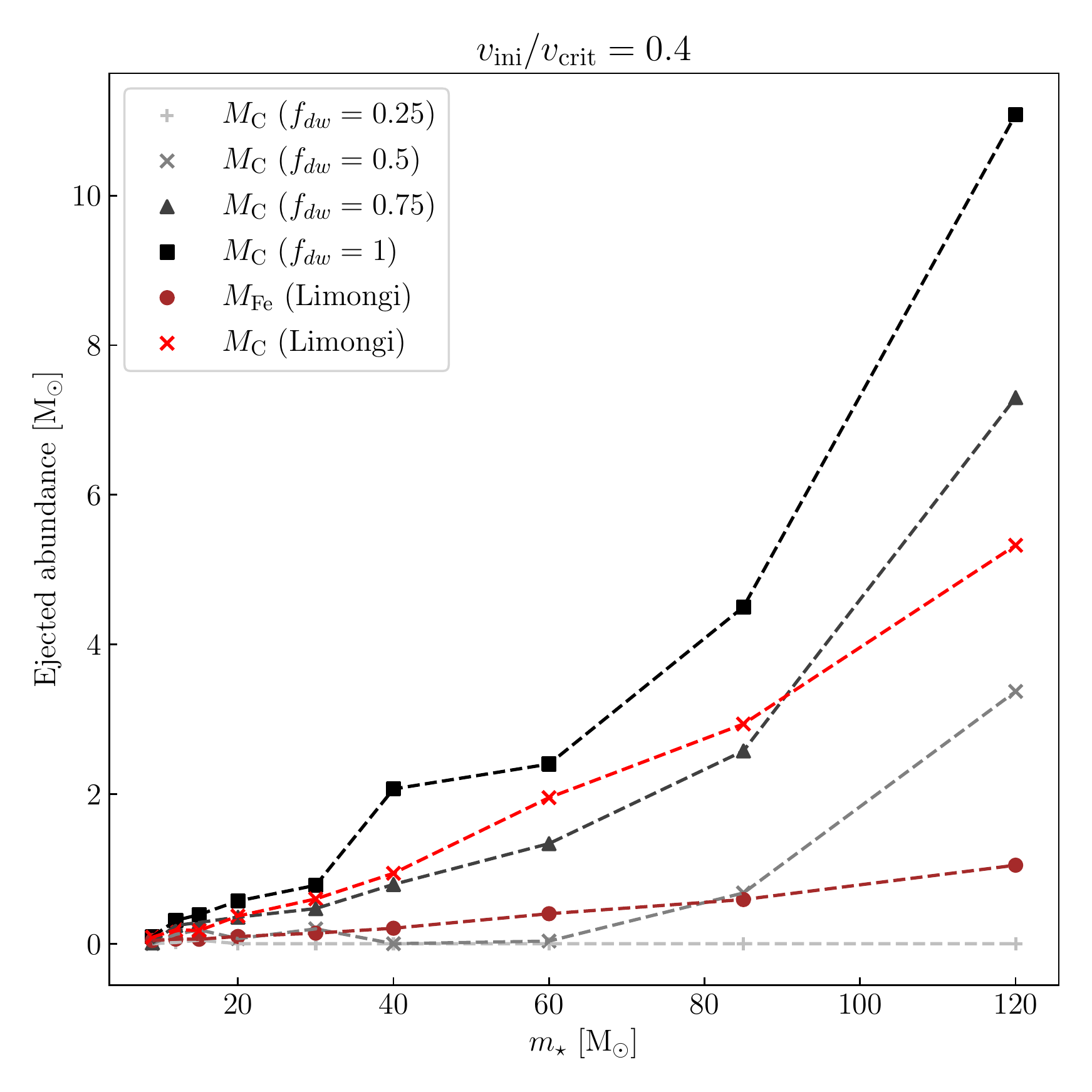}
    \vspace{-15pt}
    \caption{Ejected masses of carbon and iron as functions of the initial stellar mass and for different values of the parameter $f_{dw}$. The mass of carbon and iron ejected by the non-rotating models of \citet{Limongi2012} are also indicated. {\it Left panel:} non-rotating models ($v_{\rm ini}/v_{\rm crit}=0$). {\it Right panel:} rotating models ($v_{\rm ini}/v_{\rm crit}=0.4$).}
    \label{fig:yield}
\end{figure*}

For the mass ejected by winds, we assume that the minimum amount is the mass above the He-core. The He-core is defined as the mass inside the mass coordinate where the $^4$He mass fraction becomes larger than  0.75 starting from the surface. This is already a significant amount of mass (between 20 and 25 M$_\odot$ for a Pop III 40 M$_\odot$ star). But as we shall see below, when only that part of the star is assumed to be lost by winds, then the impact is found to be very modest. To evaluate whether Pop III winds can make a difference in primordial metal enrichment, we further consider still more extreme cases where winds go deeper into the He core. Such strong mass loss may occur from pre-SN pulsations and instabilities \citep{smith2006role,Woosley2007,Yoon2010, Fuller2017,Fuller2018,Fuller2020,Zhao2020,Fuller2021,Grassi2021}.
We adopt as the maximum amount that can be lost by winds the mass above the incipient stellar remnant. Here, by stellar remnant we mean the mass that will eventually form (out of a SN explosion) the compact object, either a neutron star or black hole, deduced from the mass of the
carbon-oxygen core obtained in the stellar grids by \citet{murphy2021grids}, and employing the relation between the mass of this core and that of the remnant from \citet{Limongi2012}. {Note that the core-remnant mass relation in \citet{Limongi2012} assumes that all stars with $m_{\star}\sim 10-120\ \rm M_{\odot}$ explode as SNe, which may not be true for Pop~III. Therefore, we refer the remnants defined by this relation as \textit{potential} stellar/SN remnants. Between these two limits, we define the wind depth parameter $f_{dw}$ as the fraction of the intermediate region from the He core mass coordinate to the \textit{potential} stellar remnant that is carried by winds. 
} 

In Fig.\ref{fig:chemie}, we show two examples of the chemical structures of the last computed models
in the grids by \citet{murphy2021grids}. They correspond to the end of the neon-photodesintegration phase for the non-rotating 40 M$_\odot$ star (left panel), and the end of the core He-burning phase for the rotating one (right panel). 
The variation of the abundance of carbon in the outer layers has already reached at those phases a pattern that is very near the one it would get at the presupernova stage. We use the 40 M$_\odot$ models as examples to illustrate Pop~III stellar evolution, since 40~M$_\odot$ is close to the typical/average stellar mass for our fiducial IMF, which follows $dN/dm_{\star}\propto m_{\star}^{-1}$ in the range of $10 - 120~\rm M_\odot$ (see Sec.~2.3). In the figure, the minimum mass assumed to be ejected by winds is shaded in grey, whereas the mass of the remnant is shown in pink. { In between, various cases are indicated by different values of $f_{dw}$.} When $f_{dw}$ is equal to 0, the grey shaded region is ejected by winds. Conversely, when $f_{dw}= 1$, the grey and the non-shaded regions are ejected, comprising all layers above the \textit{potential} stellar remnant.

We set the wind timescale to the duration of the post-main sequence (MS) phase as an upper limit, as most wind-driving mechanisms tend to occur during post-MS \citep{smith2006role,Woosley2007,Yoon2010, Fuller2017,Zhao2020,Fuller2021,Grassi2021}.
The wind velocity in turn is estimated using the expressions given by \citet[][see also Equ.~\ref{e14} in Sec.~\ref{s2.3}]{Lamers1999winds}. Note that these expressions are valid for line-driven winds, however we shall use them here for any kind of wind mass loss. Let us further note that the wind velocities are in any case proportional to the escape velocity with, in general, a rather modest proportionality factor of order 1. Figure~\ref{fig:HRD_P020z00} shows the evolution of the rotating and non-rotating 40 M$_\odot$ models in the HR diagram. We see that the rotating model evolves at lower effective temperature and higher luminosity approaching the end of the MS phase. This is a consequence of rotational mixing that produces a larger He-core at the end of the core H-burning phase, as a result of the additional rotation-induced transport of hydrogen from the envelope into the core. We have color-coded the tracks, starting from the end of the core H-burning phase, with the wind velocities given by Equ.~\ref{e14}. 

\begin{table*}
    \centering
    \caption{Carbon and iron yields from stellar winds and SNe for $f_{dw}=0.25$, 0.5, 0.75 and 1, as well as the wind velocities adopted in our metal enrichment model (see Sec.~\ref{s2.3} for details), from our model grids of non-rotating (top) and rotating (bottom) Pop~III stars. In the case of $f_{dw}$=1, winds and SNe have identical carbon yields. The iron yields from \textit{potential} SNe are derived from the CO core masses following \citet{Limongi2012}. The last stages at which chemical structure data are extracted are also shown, where $Y_{\rm c}$ denotes the core He mass fraction. See Fig.~\ref{fig:yield} for an illustration of the trends in these yields.}
    \begin{tabular}{cccccccc}
        \hline
        \hline
        $m_{\star}\ [\rm M_{\odot}]$ & $v_{w}\ [\rm km\ s^{-1}]$ &  Last stage & $M_{\rm C}\ [\rm M_{\odot}]$ & $M_{\rm C}\ [\rm M_{\odot}]$ & $M_{\rm C}\ [\rm M_{\odot}]$ & $M_{\rm C}\ [\rm M_{\odot}]$ & $M_{\rm Fe}\ [\rm M_{\odot}]$\\
        & & (Initial rotation) & ($f_{dw}=0.25$) & ($f_{dw}=0.5$) & ($f_{dw}=0.75$) & ($f_{dw}=1$, SN) & (SN)\\
        \hline
        \hline
        & & $v_{\rm ini}/v_{\rm crit}=0$\\
        \hline
        9 & 339 & Degenerate before C-ignition & 1.4e-9 & 1.9e-7 & 8.3e-5 & 0.05 & 0.04 \\
        12 & 320 & End of He-burning & 3.5e-6 & 2.7e-3 & 0.14 & 0.26 & 0.05 \\
        15 & 425 & End of He-burning & 1.8e-8 & 3.5e-5 & 0.15 & 0.37 & 0.07 \\
        20 & 215 & Ne-burning & 0.14 & 0.33 & 0.59 & 0.62 & 0.10 \\
        30 & 396 & He-burning, $Y_{\rm c}=0.007$ & 1.7e-9 & 0.31 & 0.88 & 1.42 & 0.18 \\
        40 & 310 & Ne-burning & 0.36 & 0.90 & 1.42 & 1.95 & 0.26 \\
        60 & 310 & C-burning & 0.53 & 1.15 & 1.82 & 2.57 & 0.46 \\
        85 & 382 & C-burning & 2.3e-8 & 1.65 & 2.96 & 4.27 & 0.61 \\
        120 & 352 & End of He-burning & 6.2e-8 & 1.02 & 2.34 & 3.81 & 1.01 \\
        \hline
        \hline
        & & $v_{\rm ini}/v_{\rm crit}=0.4$\\
        \hline
        9 & 276 & Degenerate before C-ignition & 6.8e-6 & 1.8e-3 & 0.01 & 0.09 & 0.04 \\
        12 & 209 & Ne-burning & 0.016 & 0.12 & 0.24 & 0.31 & 0.06 \\
        15 & 356 & C-burning & 0.048 & 0.19 & 0.28 & 0.39 & 0.06 \\
        20 & 216 & C-burning & 3.8e-5 & 0.068 & 0.36 & 0.57 & 0.10 \\
        30 & 246 & End of He-burning & 6.4e-4 & 0.20 & 0.47 & 0.78 & 0.14 \\
        40 & 328 & He-burning, $Y_{\rm c} =0.04$ & 8.4e-9 & 1.8e-5 & 0.79 & 2.07 & 0.21 \\
        60 & 262 & He-burning, $Y_{\rm c} =0.002$ & 8.2e-9 & 0.035 & 1.34 & 2.40 & 0.40 \\
        85 & 230 & He-burning, $Y_{\rm c} =0.027$ & 1.7e-8 & 0.68 & 2.58 & 4.50 & 0.59 \\
        120 & 297 & He-burning, $Y_{\rm c} =0.092$ & 2.7e-4 & 3.37 & 7.30 & 11.1 & 1.05 \\
        \hline
        \hline
    \end{tabular}
    \label{tab:yields}
\end{table*}

For the chemical composition of the wind, we have considered the abundance structures 
in the last computed models by \citet{murphy2021grids}, as shown in Fig.~\ref{fig:chemie} for the 40~M$_{\odot}$ cases. With respect to a model that would have been computed self-consistently with such a mass loss, this way of doing may overestimate the enrichment of the outer layers. In reality, those would have left the star at an earlier evolutionary stage, when any mixing induced by rotation and/or convection would not yet have changed the chemical composition of these layers. But as shown below, the enrichment is dominated by the innermost regions that will have anyway a chemical composition very similar, if not identical, to that shown in the last computed model. In Fig.~\ref{fig:chemie}, we can see the impact of rotational mixing, allowing the production of a small nitrogen overabundance (see the green dashed line in the right panel), between the mass coordinates 10 and 12.5~M$_\odot$. This nitrogen is boosted by the mixing of carbon and oxygen, produced in the He-core into the H-burning shell. This is a pocket of primary nitrogen since it is produced by carbon and oxygen newly synthesized by the star \citep{meynet2006early}.


In Table~\ref{tab:yields}, we summarize the key wind parameters. For each initial mass, we indicate the adopted wind velocity (see Sec.~\ref{s2.3}), the last computed stage of evolution, and the carbon and iron yields. For the iron yields, which are only available from SNe, we use the relation between the CO core mass and the iron yields obtained by \citet{Limongi2012}. Rotation significantly increases the mass of carbon and iron ejected by the most massive stars ($M > 85 {\rm \,M}_\odot$), while not greatly affecting the values obtained for lower initial masses. This can be seen also in Fig.~\ref{fig:yield}, where we compare our results to the SN carbon yields of \citet{Limongi2012}, which are similar to our wind yields with $f_{dw}=0.75$, for models with initial masses below 85~M$_\odot$.

\subsection{Metal enrichment model}
\label{s2.2}
Detailed modelling of metal enrichment from primordial star formation requires high-resolution radiative hydrodynamic simulations to capture the interplay between different feedback engines of massive stars, such as (ionizing) radiation, stellar winds and SN explosions (see e.g. \citealt{geen2015detailed,rahner2017winds}), which is computationally expensive and beyond the scope of this exploratory work. To efficiently investigate the mostly unconstrained parameter space of Pop~III star formation and stellar winds (in conjunction with SNe), we instead design an idealized semi-analytical model under spherical symmetry to estimate the size and mass of the wind/SN bubble carved into the primordial star-forming region, which is shaped by ionizing radiation. The bubble properties are then combined with (gravitational) turbulent metal mixing during subsequent re-collapse to determine the final dilution mass of metal yields. 

\subsubsection{Ionization feedback}
\label{s2.2.1}

\begin{figure}
    \centering
    \includegraphics[width=1\linewidth]{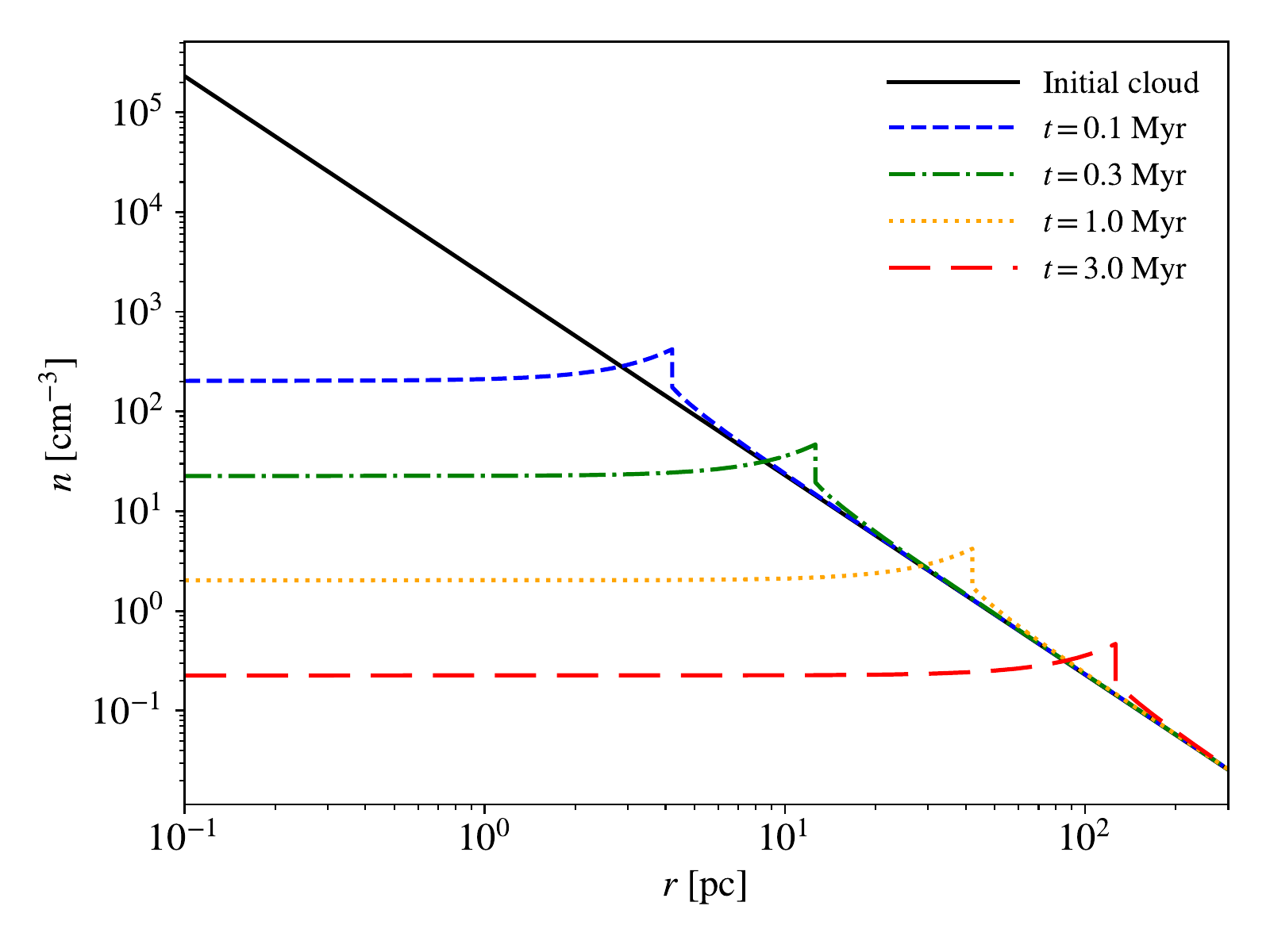}
    \vspace{-20pt}
    \caption{Density profiles from the champagne flow solution with $\epsilon\sim 0.008$ \citep{shu2002self} at $t=0.1$ (dashed), 0.3 (dashed-dotted), 1 (dotted) and 3~Myr (long-dashed), on top of the initial density profile (solid) for a typical primordial star-forming cloud.}
    \label{f1}
\end{figure}

Following \citet{alvarez2006h}, we model the ionization feedback as a D-type shock that homogenizes the downstream medium after break-out of the ionization front (I-front). The initial primordial star-forming cloud for a typical minihalo of $M_{\rm h}\sim 10^{6}\ \rm M_{\odot}$ at $z\sim 15$ can be described by a singular isothermal sphere (SIS) with a temperature of $T_{\rm SIS}\simeq 300\ \rm K$, such that the number density of hydrogen follows
\begin{align}
    n_{\rm H,SIS}(r)\simeq 2.3\times 10^{3}\ \mathrm{cm^{-3}}\left(\frac{T_{\rm SIS}}{300\ \rm K}\right)\left(\frac{r}{1\ \rm pc}\right)^{-2}\ .\label{e1}
\end{align}
The moment of break-out occurs when the I-front catches up with the shock, such that the medium enclosed by the shock is fully ionized and isothermal with a high temperature $T_{\rm\HII}\simeq 2\times 10^{4}\ \rm K$. Thereafter, the pressure imbalance across the shock front drives a `champagne' flow, resulting in homogenization and rarefaction of the downstream medium. This flow can be described by the family of self-similar solutions in \citet{shu2002self}:
\begin{align}
    x=\frac{r}{c_{s,2}t}\ ,\quad \rho(r,t)=\frac{m_{\rm H}n_{\rm H}(r,t)}{X}=\frac{\alpha(x)}{4\pi G t^{2}}\ ,\label{e2}
\end{align}
where $\rho(r,t)$ is the downstream density profile of the champagne flow whose shape is captured by $\alpha(x)$, $c_{s,2}$ is the sound speed of the ionized gas, $m_{\rm H}$ and $X=0.76$ are the mass and mass fraction of hydrogen nuclei. $\alpha(x)$ depends on the ratio $\epsilon=(c_{s,2}/c_{s,1})^{2}$, where $c_{s,1}$ is the sound speed of the upstream medium (i.e. the initial cloud). For $T_{\rm\HII}\simeq 2\times 10^{4}\ \rm K$ and $T_{\rm SIS}\simeq 300\ \rm K$, in our case $\epsilon\sim 0.008$, and the shock speed is $v_{s}=x_{s}c_{s,2}\simeq 40\ \rm km\ s^{-1}$, with $x_{s}=2.55$. Fig.~\ref{f1} shows the density profiles from the champagne flow solution at different times, where the density in the inner region ($r\ll v_{s}t$) satisfies $\rho_{\rm flat}(t)=\alpha_{0}/(4\pi G t^{2})$, with $\alpha_{0}\simeq 0.004$.

The above model is meant to capture the large-scale asymptotic behavior of the surrounding gas under ionization feedback (i.e. its long term legacy). The early stage of this dynamics remains uncertain, particularly how and when break-out happens, depending on the detailed small-scale ($r\lesssim 0.1$~pc) structure of the accretion disk at the time when stars reach the main sequence. For simplicity, we assume that the shock travels at the same speed before break-out as in the champagne flow phase in the absence of stellar winds. In this way, break-out occurs when there is enough ionization flux from the stars to fully ionize the medium behind the shock, if homogenized, leading to a break-out radius \citep{alvarez2006h}
\begin{align}
    r_{B}\simeq 2.3\ \mathrm{pc}\left(\frac{T_{\rm SIS}}{300\ \rm K}\right)^{2}\left(\frac{Q}{3\times 10^{50}\ \rm s^{-1}}\right)^{-1}\ ,\label{e3}
\end{align}
where $Q$ is the production rate of ionizing photons. This corresponds to a break-out time $t_{B}=r_{B}/v_{s}$. Since Pop~III stars are typically massive with strong ionization feedback, $Q\simeq 1.5\times 10^{51}\ {\rm s^{-1}}(M_{\star}/10^{2}\ \rm M_{\odot})$ \citep{schaerer2002properties}, break-out tends to occur early-on with $t_{B}\sim 10^{4}-10^{5}$~yr, for a typical total (initial) stellar mass of $M_{\star}\sim 100-10^{3}~\ \rm M_{\odot}$, based on simulations and observational constraints (e.g. \citealt{susa2014mass,stacy2016building,xu2016late,hirano2017formation,schauer2019constraining,danielle2020,hosokawa2020}). Considering possible slow-down of the shock before break-out due to insufficient ionization and cooling may increase the break-out time by a factor of a few, still much shorter than the lifetimes of Pop~III stars $t_{\star}\sim 3-10$~Myr. Therefore, we believe that uncertainties in the early stage have minor impact on our results for the late stage.

However, if strong stellar winds are launched before break-out, the gas is swept up to a thin shell with enhanced density that can bottle up the ionization flux, delaying the break-out (e.g. \citealt{rahner2017winds}). Such cases are expected to be rare for extremely metal-poor Pop~III stars, as line-driven winds are initially negligible (e.g. \citealt{muijres2012}) and it takes time for other wind-launching mechanisms to kick in (e.g. spinning up of the star that mixes metals into the envelope and post-MS pulsation and instabilities, see Sec.~\ref{s2.1}). Nevertheless, for completeness, we also consider this special situation with a new break-out criterion. That is when the wind bubble size reaches $r_{B}$, such that the wind shell is almost fully ionized\footnote{Here we estimate the density in the shell as the average density of the wind bubble considering all swept-up gas. The density should be higher in reality, implying a longer delay of break-out. However, we do not consider radiation pressure from the trapped ionization flux, which can facilitate bubble expansion and break-out. The two effects may cancel each other out, such that the errors here will not affect the late-time evolution that we are concerned with.}. We assume that the density profile remains the same as that of the initial cloud before break-out to calculate the wind bubble expansion (see below). We further impose an upper limit $10r_{B}/v_{s}$ to the wind-modulated break-out time $t'_{B}$ to account for the radiation pressure from the trapped ionization flux. 

\subsubsection{Dynamics of wind bubbles}
\label{s2.2.2}

The adiabatic expansion of a spherical wind bubble into a medium characteriezed by the density profile $\rho(r,t)$ under the thin-shell approximation is described by the following four equations \citep{bisnovatyi1995shock}:
\begin{align}
    \frac{dR}{dt}&=u\ ,\label{e4}\\
    \frac{dM}{dt}&=4\pi R^{2}u\rho(R,t)\ ,\label{e5}\\
    \frac{d(Mu)}{dt}&=4\pi R^{2}P\ ,\label{e6}\\
    \frac{dE}{dt}&=L_{w}(t)-4\pi R^{2}uP\ ,\label{e7}
\end{align}
where $M\equiv M_{\rm bubble}$ is the total mass of the bubble, mostly ($\gtrsim 90$\%) in the thin shell, $R\equiv r_{\rm bubble}$ is the radius of the bubble shell/rim, $u$ is the expansion speed, $E$ and $P$ are the thermal energy and pressure of the bubble, related to each other by $P=(\gamma-1)3E/(4\pi R^{3})$, given the adiabatic index $\gamma=5/3$, and $L_{w}$ is the input wind luminosity. In this picture, the vicinity of the bubble is extremely hot ($T\gtrsim 10^{6}$~K) 
with low-density. Based on the bubble structure in \citet{mac1988superbubbles} and primordial cooling model in \citet{boyuan2019}, we have verified that cooling is inefficient\footnote{We do not consider enhanced cooling by turbulent mixing at the bubble surface that may slow down bubble expansion in dense turbulent clouds \citep{coolwind2021a,coolwind2021b}, since Pop~III star formation only happens when turbulence has decayed in the initial cloud in lack of efficient cooling from metals and dust \citep{chon2021transition}, and thermal pressure can also reduce turbulence in the downstream medium of the ionization shock surrounding the wind bubble.} for $L_{w}\gtrsim 10^{36}\ \rm erg\ s^{-1}$, unless the bubble metallicity is as high as $\sim 0.1\ \rm Z_{\odot}$, which is highly unlikely for primordial clouds and Pop~III stellar winds. In all the cases explored by this work, we have $L_{w}\gtrsim 10^{36}\ \rm erg\ s^{-1}$ and $Z<0.1\ \rm Z_{\odot}$, such that the adiabatic condition (Equ.~\ref{e7}) always holds. We have ignored external thermal pressure (as well as gravity) in the momentum conservation equation~(\ref{e6}) for $t<t_{\star}$, assuming that it is balanced by radiation pressure. 

Once $L_{w}(t)$ and $\rho(r,t)$ are given, with proper initial conditions, we can solve the above equations~(\ref{e4})-(\ref{e7}) numerically to obtain the evolution of bubble variables $R$, $M$, $u$ and $E$. We stop the integration at $t=t_{\star}+t_{\rm st}$ to estimate the final size and mass of the wind bubble before re-collapse\footnote{For simplicity, we freeze the density and temperature structures at $t_{\star}$ for the bubble evolution during $t_{\star}<t< t_{\star}+t_{\rm st}$ as an 1st-order approximation.}. Here $t_{\rm st}\sim t_{\rm ff}= 4~\rm Myr$ is the timescale at which the bubble shell continues to expand after the source of winds and ionization flux has been turned off. Actually, without the heating and homogenization from ionization, the external gas cools and recombines, which then mixes rapidly into the hot bubble via the Rayleigh-Taylor instability, resulting in significant cooling that exhausts the bubble pressure \citep{mckee1984photoionized,coolwind2021a,coolwind2021b}. This happens within a few Myr, close to the free-fall timescale, $t_{\rm ff}\equiv t_{\rm ff}(r<0.1 R_{\rm vir})$, of the inner cloud in which the second generation of stars is to form given the minihalo virial radius $R_{\rm vir}\sim 200\ \rm pc$. So we use $t_{\rm ff}$ to estimate the timescale at which the bubble loses pressure and is stalled by gravity. We also require $R<r_{s}$ for $t<t_{\star}$, given $r_{s}\sim v_{s}t$ as the location of the ionization shock front, and replace $P$ with $P-c_{s}^{2}(R)\rho(R)$ at $t>t_{\star}$, taking into account the external pressure term $c_{s}^{2}(R)\rho(R)$. 

Analytical solutions of equations~(\ref{e4})-(\ref{e7}) are also available in simplified setups. For illustration, we now introduce two analytical solutions under constant $L_{w}$, which are used to set the initial conditions for numerical integration. 
First, when the surrounding medium is uniform, i.e. $\rho(R)\equiv\rm const.$, we have \citep{bisnovatyi1995shock}
\begin{align}
    R=\left[\frac{375(\gamma-1)L_{w}}{28(9\gamma-4)\pi\rho}\right]^{1/5}t^{3/5}\ .\label{e8}
\end{align}
Second, when the cloud is a SIS with $\rho(R, t)=\rho_{0}(R/r_{0})^{-2}$, equations~(\ref{e4})-(\ref{e7}) can be combined into one differential equation
\begin{align}
    \frac{d(R^{2}\ddot{R}+R\dot{R}^{2})}{dt}+3(\gamma-1)\dot{R}(R\ddot{R}+\dot{R}^{2})=\frac{3(\gamma-1)L_{w}}{4\pi\rho_{0}r_{0}^{2}}\ .\label{e9}
\end{align}
Substituting the test solution $R=At^{\xi}$ into this equation gives $\xi=1$ and
\begin{align}
    R=\left[\frac{3(\gamma-1)L_{w}}{4\pi\rho_{0}r_{0}^{2}(3\gamma-2)}\right]^{1/3}t\ .\label{e10}
\end{align}

Now, we set $t=0$ as the onset of ionization feedback and $t=t_{i}\ge 0$ as the onset of winds. If winds are launched after break-out ($t_{i}>t_{B}$), the surrounding gas has been homogenized due to ionization with $\rho(R,t)\simeq \rho(r=0,t)=\rho_{\rm flat}(t)$ given by the champagne flow solution~(\ref{e2}). In this case, we initialize the bubble with the analytical solution for a uniform medium~(\ref{e8}), fixing $\rho$ to $\rho_{\rm flat}(t_{i})$ for a small initial time step $\delta t\ll t_{\star}$. If winds start before break-out\footnote{Actually, such cases do not occur in our calculations as we only consider post-MS winds with $t_{i}\gtrsim 3\ \mathrm{Myr}\gg t_{B}$. Nevertheless, we have taken this into account for completeness.}, the bubble will first expand to the initial SIS following the analytical solution for a SIS~(\ref{e10}). The duration of this phase $t'_{B}$ is given by $R=r_{B}$ with the limit $t'_{B}\le 10t_{B}$ (see Sec.~\ref{s2.2.1} for details). The numerical integration then starts at $t_{i}+t'_{B}$, with initial conditions given by the corresponding analytical solution~(\ref{e10}). After break-out, the density profile follows $\rho(R,t)=\rho_{\rm flat}(t+t_{B}-t'_{B}-t_{i})$ with $r_{s}=v_{s}(t+t_{B}-t'_{B}-t_{i})$ ($\sim v_{s}t$ for $t\gg t_{i}+t'_{B}$) considering the delay of break-out. 
In general, for post-MS winds considered here, the wind bubble shell is always behind the ionization shock, such that ionization feedback boosts the wind bubble size but meanwhile reduce the swept-up mass as the downstream champagne flow rarefies the surrounding medium.  

\subsubsection{Supernova feedback and turbulent metal mixing}
\label{s2.2.3}

\begin{figure}
    \centering
    \includegraphics[width=1\linewidth]{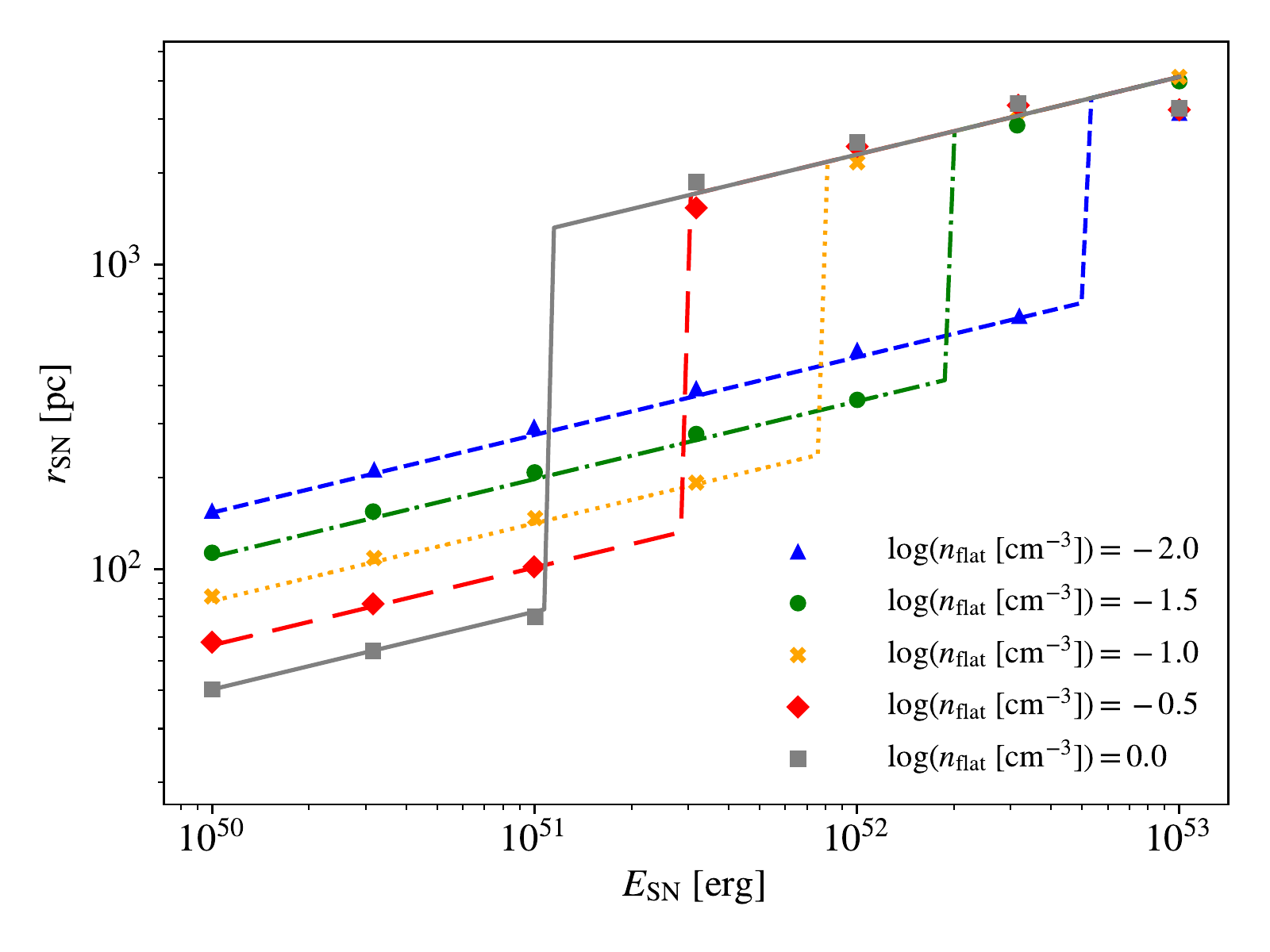}
    \vspace{-20pt}
    \caption{SN bubble radius as a function of SN energy for $n_{\rm IGM}=2\times 10^{-3}\rm cm^{-3}$ ($z\sim 20$), given $\log(n_{\rm flat}\ [\rm cm^{-3}])=-2$ (dashed, triangles), $-1.5$ (dashed-dotted, circles), $-1$ (dotted, crosses), $-0.5$ (long-dashed, diamonds) and 0 (solid, squares). Data points are from \citet[see their fig.~4]{ji2015preserving}, fitted by our formula~(\ref{e11}), shown with curves of different styles.}
    \label{f3}
\end{figure}

The wind bubble mass is not the final dilution mass $M_{\rm dilution}$ of metal enrichment, as SN explosions that may sweep up more materials and turbulent mixing during re-collapse can further boost $M_{\rm dilution}$. Here we follow the ideas in \citet{ji2015preserving} to take into account the relevant effects. 

We estimate the radius of the SN bubble $r_{\rm SN}$ by fitting the numerical results in \citet[see their fig.~4]{ji2015preserving} with a power-law formula, taking into account the fact that the SN bubble can expand much further into the intergalactic medium (IGM) once it goes beyond the ionization shock:
\begin{align}
    r_{\rm SN}=\begin{cases}
    r_{\rm SN,d}\ ,\quad r_{\rm SN,d}\le 1.25 r_{s}(t_{\rm SN}) \ ,\\
    r_{\rm SN,u}\ ,\quad r_{\rm SN,d}>1.25 r_{s}(t_{\rm SN}) \ ,\\
    \end{cases}\label{e11}
\end{align}
where $r_{s}(t_{\rm SN})$ is the radius of the ionization shock front when the SN explosion occurs at $t=t_{\rm SN}$, and
\begin{align}
\begin{split}
     r_{\rm SN,d}&\simeq 40\ {\rm pc}\cdot \left(\frac{E_{\rm SN}}{10^{50}\ {\rm erg}}\right)^{b_{1}}\left(\frac{n_{\rm flat}}{1\ {\rm cm^{-3}}}\right)^{b_{2}}\ ,\\\
     r_{\rm SN,u}&\simeq 4.1\ {\rm kpc} \left(\frac{E_{\rm SN}}{10^{53}\ {\rm erg}}\right)^{b_{1}}\left(\frac{n_{\rm IGM}}{2\times 10^{-3}\rm cm^{-3}}\right)^{b_{2}}\ .
\end{split}\label{e12}
\end{align}
Here $E_{\rm SN}$ is the SN energy, $b_{1}=0.255$, $b_{2}=-0.29$, $n_{\rm flat}=\rho_{\rm flat}(t_{\rm SN})X/m_{\rm H}$ is the central density in the shocked region, and $n_{\rm IGM}=\bar{\rho}_{m}(z)\Omega_{b}/\Omega_{m}\simeq 8\times 10^{-4}\ {\rm cm^{-3}}[(1+z)/16]^{3}$ is the density in the IGM. Fig.~\ref{f3} shows how our formula~(\ref{e11}) fits the results in \citet{ji2015preserving} for $n_{\rm IGM}=2\times 10^{-3}\rm cm^{-3}$ (i.e. $z\sim 20$). When SN explosions are present, we set the initial enrichment/dilution radius to $r_{\rm shell}=\max(r_{\rm SN},r_{\rm bubble})$ for simplicity\footnote{In reality, the interaction between SN ejecta and wind bubbles can be complex. We adopt $r_{\rm shell}=\max(r_{\rm SN},r_{\rm bubble})$ as a approximation to the final momentum-driven phase for the combined SN-wind shell.}.

Given $r_{\rm shell}$, we still need to model turbulent mixing during re-collapse. For simplicity, we follow \citet{ji2015preserving,karlsson2008uncovering} to work in the Lagrangian picture, where we calculate a mixing radius for the snapshot before re-collapse, assuming that metals will be fully mixed into the medium initially enclosed by this radius. In this model, the turbulent mixing is driven by gravity during re-collapse, such that the mixing radius can be estimated as \citep{ji2015preserving,karlsson2008uncovering}
\begin{align}
    r_{\rm mix}=(6D_{\rm t}t_{\rm mix}+r_{\rm shell}^{2})^{1/2}\ ,\label{e13}
\end{align}
where $t_{\rm mix}$ is the timescale of mixing/re-collapse, and $D_{\rm t}=\langle v_{\rm turb}\rangle\langle l_{\rm turb}\rangle/3\simeq\beta V_{\rm vir}R_{\rm vir}/3$ is the turbulent diffusion coefficient characterized by the dimensionless parameter $\beta$, the virial velocity $V_{\rm vir}$ and radius $R_{\rm vir}$ of the host halo of the \textit{second} generation of stars. The final dilution mass is then given by the total mass enclosed within $r_{\rm dilution}=r_{\rm mix}$. Throughout this work, we set the turbulent mixing parameter $\beta=0.15$, which implies halo-scale mixing for atomic-cooling (AC) haloes according to the cosmological simulations in \citet{liu2020did}. 

\subsubsection{Applications to typical high-$z$ host haloes}
\label{s2.2.4}

\begin{figure}
    \centering
    \includegraphics[width=1\linewidth]{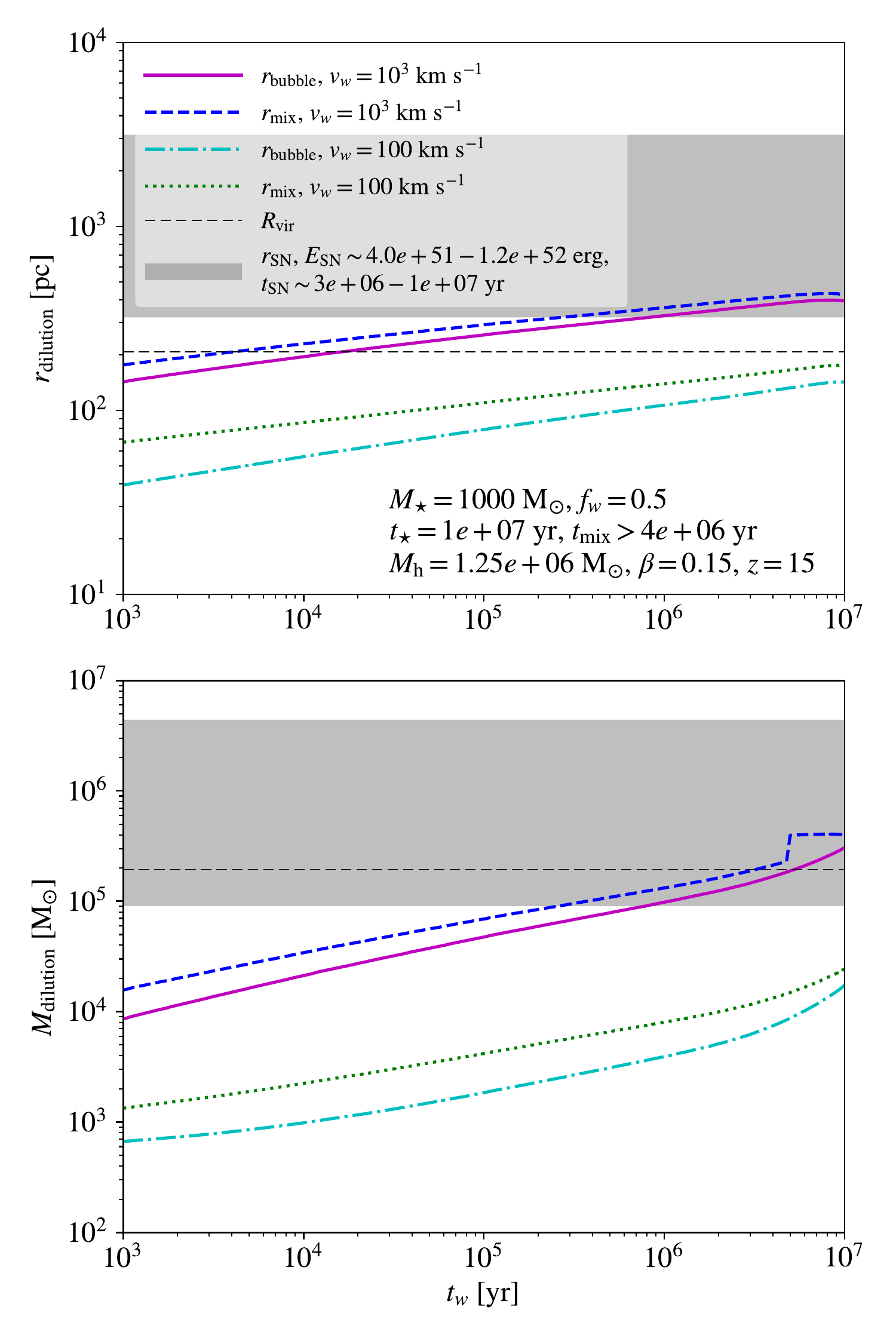}
    \vspace{-20pt}
    \caption{Dilution radius (top) and mass (bottom) as functions of wind timescale $t_{w}$ for metal enrichment from a Pop~III cluster of $M_{\star}= 10^{3}\ \rm M_{\odot}$ in a minihalo of $M_{\rm h}=1.25\times 10^{6}\ \rm M_{\odot}$ at $z= 15$. The results for wind enrichment without SN explosions are shown with curves of the normal line width and different styles, for $v_{w}\sim 100-10^{3}\ \rm km\ s^{-1}$, before (solid and dashed-dotted) and after (dashed and dotted) turbulent mixing. For comparison, the shaded regions show the results for SN bubbles with $E_{\rm SN}=0.4-1.2\times 10^{52}\ \rm erg$ at $t_{\rm SN}=3-10\ \rm Myr$. The thin horizontal dashed lines denote the properties of the gas content of the halo within $R_{\rm vir}$ (ignoring the ionization shock).}
    \label{f4}
\end{figure}

We now apply the above formalism to two situations. In the first case, as our main focus, we consider an isolated minihalo of $M_{\rm h}=1.25\times 10^{6}\ \rm M_{\odot}$ at $z=z_{1}= 15$ {(with a virial temperature of $T_{\rm vir}\simeq 2000\ \rm K$)} subject to self-enrichment. {We choose such specific mass and redshift to capture the typical sites of Pop~III self-enrichment. Due to Lyman-Werner feedback, Pop~III star formation is shifted to more massive haloes at lower redshifts. According to the cosmological simulation in \citet{liu2020did}, the Pop~III star formation rate in minihaloes is almost identical to that in AC haloes at $z\sim 15$, such that the majority of self-enrichment events happens at $z\gtrsim 15$. Since the Pop~III star formation rate density increases in time until $z\sim 10$, we chose $z=15$ as the typical formation redshift of self-enriched minihaloes. Actually, our results are insensitive to the minihalo mass and redshift, given their indirect influence via the IGM density $n_{\rm IGM}$ (Equ.~\ref{e12}) and turbulent diffusion coefficient $D_{\rm t}$ (Equ.~\ref{e13}). The former is only relevant when the SN bubble breaks out into the IGM, which is rare in our case (see Sec.~\ref{s3.1}). While for the latter, the turbulent diffusion term is rather unimportant in determining the size/mass of the metal-mixing region (see Fig.~\ref{f4}).}

The re-collapse timescale is set to $t_{\rm mix}=\max[t_{\rm ff}(<r_{\rm shell}) ,4\ \rm Myr]$, where $t_{\rm ff}(<r_{\rm shell})$ is the free-fall timescale\footnote{We have checked that the cooling/recombination timescale is smaller than the free-fall timescale.} of the feedback bubble, and the lower limit $4$~Myr is the free-fall timescale of the inner region with $r<0.1R_{\rm vir}$ in the initial cloud, which contains enough gas to form stars \citep{bromm2002}. We also impose an upper limit of $100$~Myr on $t_{\rm mix}$, which is the typical timescale for such minihaloes to merge into larger structures (see the second case below). For illustration, we now consider a Pop~III cluster of $M_{\star}= 10^{3}\ \rm M_{\odot}$, assuming that half of the initial stellar mass is lost by winds at a constant rate in a timescale of $t_{w}$ during the late stage of stellar evolution with a lifetime of $t_{\star}=10$~Myr. That is to say, $\dot{M}_{w}=0.5M_{\star}/t_{w}$ and $t_{i}=t_{\star}-t_{w}$. The resulting dilution radius and mass as functions of $t_{w}$ are shown in Fig.~\ref{f4} for $v_{w}\sim 100-10^{3}\ \rm km\ s^{-1}$ and $t_{w}\sim 10^{-3}-10\ \rm Myr$. 

It turns out that the dilution mass of wind enrichment in the absence of SNe generally follows $M_{\rm dilution}\propto t_{w}^{1/3} v_{w}^{6/5}$. The reason is that for post-MS winds at a timescale much smaller than that of the ionization feedback, the density of external medium evolves slowly, such that Equ.~\ref{e8} holds approximately, leading to $M_{\rm dilution}\propto r_{\rm bubble}^{3}\propto L_{w}^{3/5}\propto v_{w}^{6/5}$. Besides, with smaller $t_{w}$, the bubble expands at a larger rate with enhanced energy loss when the star is alive (see Equs.~\ref{e6} and \ref{e7}). As a result, the final bubble size before re-collapse is reduced. 

If SN explosions are present, the initial enrichment radius can be larger, $r_{\rm SN}\sim$ a few hundred pc, such that the re-collapse timescale is also longer, $t_{\rm mix}\sim t_{\rm ff}(<r_{\rm SN}) \sim 100$~Myr. As SNe are energetic events with energies $E_{\rm SN}\gtrsim 10^{51}\ \rm erg$, in most cases turbulent mixing only increases the dilution radius by $\lesssim 20$\%, and we have $M_{\rm dilution}\sim M_{\rm SN}$, where $M_{\rm SN}$ is the gas mass enclosed within $r_{\rm SN}$. In Fig.~\ref{f4} we also show the results for SN bubbles from one-time\footnote{Since the SN bubble expansion timescale $\sim 100$~Myr is much longer than the lifetimes of massive Pop~III stars $\sim 3-10$~Myr, we approximate a series of explosions with one final explosion by adding up the energies of individual SNe for simplicity.} explosions with SN energies 
$E_{\rm SN}=0.4-1.2\times 10^{52}\ \rm erg$ at $t_{\rm SN}=3-10\ \rm Myr$. For strong winds with $v_{w}\sim 10^{3}\ \rm km\ s^{-1}$, $t_{w}\gtrsim 2$~Myr and SN explosions of $E_{\rm SN}\gtrsim 4\times 10^{51}\ \rm erg$, the initial dilution radius already exceeds the halo virial radius $R_{\rm vir}\sim 200$~pc, whereas turbulent mixing is unimportant. 

In the second case, we consider an AC halo of $M_{\rm h}=10^{8}\ \rm M_{\odot}$ at $z=z_{2}= 10$, whose metal content is expected to come from multiple minihaloes with previous star formation that merge in a timescale of $t_{\rm merge}\sim t_{\rm mix}\sim 100$~Myr. For simplicity, we do not model this process in detail (e.g. with merger trees), but capture it with effective parameters. As the AC halo is more massive with stronger gravity and longer assembly timescales compared with minihaloes, containing the contributions of multiple stellar systems, turbulent mixing is the dominant mechanism, such that the initial (potentially distinct) feedback features encoded in $r_{\rm shell}$ for individual clusters are reduced. Therefore, we set the \textit{effective} mixing radius as $r_{\rm mix}\sim (6D_{\rm t}t_{\rm mix})^{1/2}\sim 850$~pc for $\beta=0.15$. Note that here we only consider the metal enrichment of the central metal mixing region, ignoring satellites (which can be similar to self-enriched minihaloes).

Different from the case of a self-enriched minihalo where the source of metals is at the halo center, metals can be carried into a AC halo from progenitors of different orbits, such that the \textit{effective} mixed volume may not be concentrated around the center. In light of this, we estimate the dilution mass as $M_{\rm dilution}\sim M_{\rm gas}(r_{\rm mix}/R_{\rm vir})^{3}-M_{\rm gas}r_{\rm mix}/R_{\rm vir}\sim 4.3-10\times 10^{6}\ \rm M_{\odot}$, given the virial radius $R_{\rm vir}\sim 1.3$~kpc and the total gas mass $M_{\rm gas}\sim M_{\rm h}\Omega_{b}/\Omega_{m}$ of the halo, which covers the range of concentration of the mixed gas. Here the lower limit is for a uniform distribution, while the upper limit corresponds to a SIS, which is a good approximation to the gas density profile in high-$z$ AC haloes from cosmological simulations (e.g. \citealt{shang2010supermassive,regan2014numerical,wise2019formation}).

Finally, we define $N_{\rm PopIII}$ as the total number of Pop~III clusters that fall into the AC halo. We will consider three cases with $N_{\rm PopIII}=10$, 30 and 100, 
which is meant to cover the range of assembly histories. It is possible that only a fraction of the yields from the Pop~III cluster falls into the central mixing region, especially for clusters with the strongest feedback ($r_{\rm shell}\gtrsim R_{\rm vir}$). To take this into account, we follow \citet{ji2015preserving} to multiply the contribution of each cluster 
by an overlapping factor $f_{\rm over}=V(r,r_{\rm shell}, Ar_{\rm mix})/[4\pi (r_{\rm shell})^{3}/3]$, given $V(r,r_{\rm shell}, Ar_{\rm mix})$ as the overlapping volume of two spheres of radii $r_{\rm shell}$ and $ Ar_{\rm mix}$ at a distance $r$, drawn randomly for a uniform (density) distribution ($p(r)\propto r^{2}$) or a SIS ($p(r)=\rm const.$) within $r<A R_{\rm vir}$. Here $A=(\Delta)^{1/3}(1+z_{2})/(1+z_{1})$ is a parameter that captures the evolution of the Lagrangian radius, where we set the overdesnity $\Delta=200$ for both the mixing region and the halo as a whole. $z_{1}$ is the redshift at which the Pop~III cluster is formed, generated randomly from a distribution following the average number of star-forming progenitors\footnote{In our case, star-forming progenitors are defined as haloes with $M_{1}>M_{\rm SF}(z_{1})$, where $M_{\rm SF}$ is the mass threshold for efficient $\rm H_{2}$ cooling. We adopt the $M_{\rm SF}$ formula from \citet{trenti2009formation}, and further impose a lower limit of $10^{6}\ \rm M_{\odot}$ based on the simulations of \citet{Anna2018} for the typical case of $1\sigma$ baryon-dark-matter streaming velocity.} as a function of redshift, $p(z_{1})\propto N_{\rm pro}(z_{1})$ in the range of $z_{2}<z_{1}<20$. We calculate $N_{\rm pro}$ by integrating the progenitor mass function obtained with the standard EPS formalism \citep[see \citealt{liu2020xlssc122} for details]{mo2010galaxy}. Note that stars formed in AC haloes can also be contaminated with metals from Pop~II stars, which can easily erase Pop~III signatures {\citep{ji2015preserving,salvadori2019}}. We defer the investigation for this effect to future work with more detailed treatments of structure formation and metal mixing. 

\subsection{Pop~III star formation model}
\label{s2.3}

\begin{table*}
{
    \caption{Key model parameters. The upper panel shows the adjustable parameters for Pop~III star clusters, wind and SN properties, as well as AC haloes, which are explored in Sec.~\ref{s3}. Here $v_{\rm crit}$ is the break-up velocity at critical rotation (see equs.~3 and 4 in \citealt{hirschi2007very}), and $M_{\rm PISN}^{\min}=120\ \rm M_{\odot}$ is the minimum mass for pair-instability SNe \citep{tanikawa2020fitting}. The bottom panel shows the stochastic parameters for fallback-mixing in faint SNe and ISM accretion, whose distributions are chosen to reproduce the observed distribution of $\rm [Fe/H]$ for CEMP-no stars in the fiducial model for self-enriched minihaloes (defined in Sec.~\ref{s3.1.2}).}
    \centering
    \begin{tabular}{cccc}
    \hline
        Parameter & Definition (\& probability distribution) & Range & Fiducial \\
    \hline
        $v_{\rm ini}$ & Initial equatorial stellar rotation velocity & 0 or $0.4 v_{\rm crit}$ & $0.4v_{\rm crit}$\\
        $f_{dw}$ & Wind depth parameter: fraction of the intermediate region lost in winds & $0-1$ & 0.5 \\
        & between the He core mass coordinate and the \textit{potential} stellar remnant\\
        $m_{\rm CCSN}^{\max}$ & Upper limit of stellar mass for CCSNe & $25\ \mathrm{M_{\odot}}$ or $M_{\rm PISN}^{\min}$ & $25\ \rm M_{\odot}$\\
        $\alpha$ & Power-law slope of the IMF $dN/dm_{\star}\propto m_{\star}^{-\alpha}$ & $0-2.35$ & 1\\
        SFE & Star formation effciency in terms of the total stellar mass $M_{\star}$ & $\sim 80-900\ \rm M_{\odot}$ & $\sim 150\ \rm M_{\odot}$\\
        $N_{\rm pro}$ & Number of Pop~III star forming progenitors of an AC halo & $10-100$ & -\\
        $p(r) $ & Distribution of progenitors in AC haloes: $p(r)\equiv 4\pi r^{2}\rho(r)$ & $\propto r^{2}$ or constant & - \\
    \hline
        $f_{\rm Fe,ej}$ & Reduction factor of the iron yield by fallback-mixing in faint SNe & $10^{-5}-1$ & - \\
        & $p(f_{\rm Fe,ej})\propto f_{\rm Fe,ej}^{\alpha_{\rm fSN}}$ with $\alpha_{\rm fSN}=-1/2$ in the range $[10^{-5},1]$ & \\
        $\Delta Y_{\rm Fe}$ & Increase of Iron abundance by ISM accretion in units of the solar value & $0-5\times 10^{-4}$ & -\\
        & $p(\Delta Y_{\rm Fe})\propto\Delta Y_{\rm Fe}^{\alpha_{\rm ISM}}$ with $\alpha_{\rm ISM}=-0.4$ in the range $[10^{-8},5\times 10^{-4}]$ \\
    \hline
    \end{tabular}
    \label{tab:para}
}
\end{table*}

{We now describe our representation of Pop~III star formation, providing stellar feedback inputs to the above metal enrichment model, with key parameters summarized in Table~\ref{tab:para}}. Recent advancements in hydrodynamic simulations of primordial star-forming clouds have shown that fragmentation is also present in extremely metal-poor/free environments and Pop~III stars are likely formed in small clusters of a few members (see e.g. \citealt{haemmerle2020formation,shima2021}). In light of this, we take Pop~III clusters as the basic units of Pop~III formation and source of early metal enrichment, rather than individual stars. Throughout our derivation of feedback parameters, we have ignored the effects of binaries, since close binaries (with separations $a\lesssim 100$~AU) with effective binary interactions are likely rare ($\lesssim 1\%$) in Pop~III systems \citep{liu2021dynamical}. 

We consider two approaches of constructing the cluster, which denote different levels and scatters in Pop~III star formation efficiency (SFE). { Here we assume that the initial star-forming clouds of Pop~III clusters have similar masses, such that the SFE by our definition is determined by the total stellar mass $M_{\star}$ of a Pop~III cluster. }

In Approach I, the total number of stars is pre-determined, which is drawn from a uniform distribution in $N_{\star}\sim 1-N_{\rm max} $, where $N_{\rm max}$ is an adjustable parameter. The masses of individual stars $m_{\star,j}$ are then generated by sampling an input IMF\footnote{Linear interpolation over (initial) stellar mass is used to derive all stellar properties from the stellar evolution grids.}. We consider two cases with $N_{\max}=6$ and 12 (corresponding to average total stellar masses $\langle M_{\star}\rangle\sim 150-300\ \rm M_{\odot}$ for the fiducial IMF defined below). In Approach II, the total stellar mass $M_{\star}$ is fixed, and the input IMF is sampled until we have $\sum_{j}m_{\star,j}\ge M_{\star}$ from individual stars\footnote{We cap the mass of the last star to enforce $\sum_{j}m_{\star,j}=M_{\star}$.}. We consider two mass values, $M_{\star}\simeq 400$ and $900\ \rm M_{\odot}$, higher than those seen in the above cases of Approach I, based on simulations and observational constraints that give $M_{\star}\gtrsim 500\ \rm M_{\odot}$ (e.g. \citealt{susa2014mass,stacy2016building,xu2016late,hirano2017formation,schauer2019constraining,danielle2020,hosokawa2020}). 

{ For the IMF, we fixed the range of (initial) stellar mass to $m_{\star}\sim 10-120\ \rm M_{\odot}$ throughout the calculation and assume a power-law shape $dN/dm_{\star}\propto m_{\star}^{-\alpha}$ for simplicity. We explore three cases for the IMF slope: $\alpha=1$ (fiducial), $0$ (top-heavy) and $2.35$ (bottom-heavy). }

Given a Pop~III cluster, we set the overall lifetime of ionizing sources to $t_{\star}=\max\{t_{\star,j}\}$, where $t_{\star,j}$ are the lifetimes of individual stars. Since the break-out time, $t_{B}$, of the ionization shock is much smaller than the lifetimes of individual stars, in the calculation of $t_{B}$ (see Equ.~\ref{e3}), we take into account all stars and set the ionization flux to $Q\simeq 1.5\times 10^{51}\ {\rm s^{-1}}(M_{\star}/10^{2}\ \rm M_{\odot})$ \citep{schaerer2002properties}.

For SN feedback, we combine all SNe into one explosion of $E_{\rm SN}$ by adding up their energies: $E_{\rm SN}=\sum_{j}E_{{\rm SN},j}$. The onset of this explosion is set to $t_{\rm SN}=\sum_{j}E_{\mathrm{SN},j} t_{\star,j}/E_{\rm SN}$. For each SN, the explosion energy $E_{{\rm SN}, j}$ is generated randomly from a log-flat distribution in the range of $E_{{\rm SN},i}\sim 0.3-10\times 10^{51}\ \rm erg$. We only consider core-collapse SNe (CCSNe) from stars in the mass range of $m_{\star}\sim 10-120\ \rm M_{\odot}$ (covered by our stellar evolution models), and defer the consideration of (pulsational) pair-instability SNe (PISNe) to future studies\footnote{For simplicity, we ignore the effect of rotation on the final fates of Pop~III stars, which can be significant for $v_{\rm ini}/v_{\rm crit}\gtrsim 0.5$ \citep{yoon2012evolution}.} {(see e.g. \citealt{debennassuti2017,salvadori2019})}. In the fiducial case (without special annotation or `fiducial SNe'), only stars with $m_{\star}<M_{\rm CCSN}^{\max}=25\ \rm M_{\odot}$ will explode, while more massive stars will collapse into black holes (BHs) directly with no ejecta. { Here we adopt the upper limit $M_{\rm CCSN}^{\max}=25\ \rm M_{\odot}$ predicted by the stellar evolution models in \citet[see their equs.~51-53]{tanikawa2020fitting} as a conservative choice. We also consider another extreme case in which all stars end up in CCSNe, denoted by `all SNe', i.e. $M_{\rm CCSN}^{\max}=M_{\rm PISN}^{\min}$ where $M_{\rm PISN}^{\min}=120\ \rm M_{\odot}$ is the minimum mass for PISNe \citep{tanikawa2020fitting}.} 

We further investigate the scenario of faint SNe. Similar to the idealized treatment in \citet{komiya2020faint}, to capture the fallback-mixing effect, we multiply the iron yield by a factor $f_{\rm Fe,ej}$, randomly generated from a power-law distribution $p(f_{\rm Fe,ej})\propto f_{\rm Fe,ej}^{\alpha_{\rm fSN}}$ with $\alpha_{\rm fSN}=-1/2$ in the range $[10^{-5},1]$ for each SN, leaving the yields of other elements unchanged. The distribution of $f_{\rm Fe,ej}$ is chosen to reproduce the shape of the observed distribution of $\rm [Fe/H]$ for CEMP-no stars at $[\rm Fe/H]\lesssim -4$ in the fiducial model {(defined in Sec.~\ref{s3.1.2})} for self-enriched minihaloes (see below). 

Finally, for stellar winds, we again combine all stars with potentially multiple pulses of winds into one episode for simplicity. The onset of this episode is set to the minimum MS lifetime for all stars (with $m_{\star}\ge 10\ \rm M_{\odot}$), $t_{i}=\min\{t_{{\rm MS},j}\}$, since strong winds (triggered by rotation) tend to occur in the post-MS stage (see e.g. \citealt{smith2006role,meynet2006early,hirschi2007very,ekstrom2008effects}). As an upper limit, the duration of this pulse is the average of the post-MS timescale of individual stars $t_{{\rm PMS},j}$, weighted by the wind energy $E_{w,j}=0.5M_{w,j}v_{w,j}^{2}$, i.e. $t_{w}=\sum_{j} t_{{\rm PMS},j}E_{w,j}/(\sum_{j} E_{w,j})$. Here we have $t_{{\rm PMS},j}=0.1 t_{{\rm MS},j}$ as an approximation, and $M_{w,j}$ and $v_{w,j}$ are the mass lost in winds and wind velocity for star $j$. Considering smaller $t_{w}$ and/or $v_{w}$ will reduce the dilution mass of wind enrichment and thus increase carbon enhancement, as $M_{\rm dilution}\propto t_{w}^{1/3}v_{w}^{6/5}$ approximately (see Sec.~\ref{s2.2.4}). The wind luminosity during $t_{i}<t<t_{i}+t_{w}$ is then given by $L_{w}=\sum_{j} E_{w,j}/t_{w}$. 

Since our stellar evolution models do not include strong winds explicitly, we treat the wind mass loss $M_{w,j}$ as a free parameter. For simplicity, we further relate $M_{w,j}$ to the wind depth parameter $f_{dw}$ as the fraction of mass between the 75\% He core\footnote{The mass coordinate at which the $\rm ^{4}He$ abundance reaches 75\%.} and the \textit{potential} SN remnant carried by winds (see Sec.~\ref{s2.1}). Here we have assumed that all mass above the He core is lost in winds, and $f_{dw}$ is independent of stellar mass. Four cases with $f_{dw}=0.25$, 0.5, 0.75 and 1 are considered, where $f_{dw}=0.5$ is regarded as the fiducial case. For each model in our grids, the wind velocity $v_{w}$ is computed using scaling relations with the escape velocity $v_{\rm esc}$ from \citet[pp. 47 \& 53]{Lamers1999winds}:
\begin{align} 
    v_{w,j}= 
    \begin{cases}
    \frac{v_{\rm esc}}{2.6}\ ,\quad T_{\rm eff}>21000~\rm K\\
    \frac{v_{\rm esc}}{1.3}\ ,\quad 1000~\mathrm {K}<T_{\rm eff}<21000~\rm K\\
    \frac{v_{\rm esc}}{0.7}\ ,\quad  8000~\mathrm {K}<T_{\rm eff}<10000~\rm K\\
    \frac{v_{\rm esc}}{0.5}\ ,\quad T_{\rm eff}<8000~\rm K
    \end{cases}\label{e14}
\end{align}
We compute $v_{\rm esc}$ at each timestep of the post-MS evolution\footnote{
Since significant mass loss is not explicitly included in our stellar evolution grids, we may underestimate the escape velocity in the last phases of the evolution when the star tends to become more compact after strong mass loss.}, and the wind velocity is then derived from Equ.~(\ref{e14}). As a conservative estimate, i.e. lower limit\footnote{As $M_{\rm dilution}\propto t_{w}^{0.33} v_{w}^{1.2}$ approximately (see Sec.~\ref{s2.2.4}), $t_{w}$ and $v_{w}$ are degenerate parameters for wind feedback. While we have adopted an upper limit for $t_{w}$, here we instead consider a lower limit for $v_{w}$ by taking the value close to the end of evolution, as $v_{w}$ tends to decrease with time during the post-MS phase (see Fig.~\ref{fig:HRD_P020z00}). In this way, we expect to capture the `median' of wind properties over the uncertain parameter space of Pop~III winds.}, we assign a single value of $v_{w}$ by adopting 
the average wind velocity over the last stage of evolution with a duration of $10^{-3}t_{\mathrm{ MS},j}$. 
In this way, we have $v_{w}\sim 200-400\ \rm km\ s^{-1}$ for $m_{\star}\sim 10-120\ \rm M_{\odot}$. The typical wind luminosity from a Pop~III cluster of $M_{\star}\sim 200\ \rm M_{\odot}$ is $L_{w}\sim 10^{37}\ \rm erg\ s^{-1}$ (with $t_{w}\sim 0.5\ \rm Myr$, $\dot{M}_{w}\sim 3\times 10^{-4}\ \rm M_{\odot}\ yr^{-1}$, $v_{w}\sim 300\ \rm km\ s^{-1}$).

Once the feedback parameters ($t_{\star}$, $t_{\rm SN}$, $E_{\rm SN}$, $t_{i}$, $t_{w}$, $L_{w}$) are known as described above for a given halo of mass $M_{\rm h}$ at redshift $z$, we can derive the dilution mass $M_{\rm dilution}$ and size $r_{\rm mix}$ of the metal mixing region. For a self-enriched minihalo with a single Pop~III cluster, the mass fraction of element $k$ in the enriched cloud is then given by
\begin{align}
    Y_{k}=M_{k}/M_{\rm dilution}=\sum_{j}(m^{{\rm SN},j}_{k}+m^{w,j}_{k})/M_{\rm dilution}\ ,
\end{align}
where $M_{k}$ is the total yield of $k$ from the contributions of SNe and winds for individual stars, $m^{{\rm SN},j}_{k}$ and $m^{w,j}_{k}$. While for an AC halo enriched by multiple Pop~III clusters, we have
\begin{align}
    Y_{k}=\sum_{l}M_{k,l}f_{{\rm over},l}/M_{\rm dilution}\ ,
\end{align}
where $M_{k,l}$ is the yield from cluster $l$, and $f_{{\rm over},l}\equiv f_{{\rm over}}(r_{l},r_{{\rm shell},l},A_{l}r_{\rm mix})$ is the overlapping factor defined in Sec.~\ref{s2.2.4}. Here the (central) mixing region size $r_{\rm mix}$ and mass $M_{\rm dilution}$ are determined by turbulent mixing of the AC halo itself, independent of the feedback bubbles of individual Pop~III clusters that fall into the AC halo. 

We also consider the effect of surface pollution by accretion from the ISM with post-processing. The ISM accretion process has long been proposed to explain the absence of Pop~III survivors and extremely low metallicities of EMP stars in observations (e.g. \citealt{yoshii1981,iben1983,Shigeyama2003,Komiya2009,frebel2009minimum,komiya2015most}). 

{
It is found in structure formation models (e.g. \citealt{komiya2015most}) that the surface iron abundance obtained through ISM accretion spans a wide range of $-8\lesssim [\rm Fe/H]\lesssim -2$, reflecting the diversity of merger histories. For simplicity, we do not model this process explicitly but calibrate our model to observational results with a simple approach. We add $\Delta Y_{\rm Fe}$ to the (surface) iron abundance\footnote{The abundances of other elements are increased accordingly assuming the solar abundance pattern for the ISM.} of each system, where $\Delta Y_{\rm Fe}$ (in units of the solar value) is randomly drawn from a power-law distribution $p(\Delta Y_{\rm Fe})\propto\Delta Y_{\rm Fe}^{\alpha_{\rm ISM}}$ with $\alpha_{\rm ISM}=-0.4$ in the range $[10^{-8},5\times 10^{-4}]$. The distribution of $\Delta Y_{\rm Fe}$ is chosen to match the observed distribution of $\rm [Fe/H]$ for CEMP-no stars (at $[\rm Fe/H]\lesssim -4$) in our fiducial model for self-enriched minihaloes {(defined in Sec.~\ref{s3.1.2})} for self-enriched minihaloes (see below), which is also within the range covered by theoretical predictions \citep{komiya2015most}.}

\section{Chemical signatures of primordial metal enrichment}
\label{s3}

In this section we present the chemical signatures of second-generation stars predicted by our metal enrichment model. All abundances and abundance ratios are shown in solar units (ignoring the slight difference between primordial and solar He abundance). {We systematically explore a wide range of situations
, considering uncertainties in Pop~III wind and SN yields, IMF and SFE, as well as AC halo modelling (see Tabel~\ref{tab:para} for the range and fiducial values of the relevant free parameters).} 
The key features of our predictions and their dependence on model parameters are demonstrated for self-enriched minihaloes (Sec.~\ref{s3.1}) and AC haloes (Sec.~\ref{s3.2}), {in comparison with previous theoretical models \citep{hartwig2019formation,komiya2020faint,jeon2021role} and observations \citep{cooke2014carbon,placco2014carbon,yoon2016observational}.} We focus particularly on the effects of winds, and reproducing the observed population of CEMP-no stars as the bona-fide footprints of primordial metal enrichment. For a given halo model, we assume that CEMP-no stars in the MW come from multiple such haloes, each contributing the same number of second-generation stars, as a zeroth-order approximation to the complex assembly history of the MW. 

{
Note that there are more recent observational data for the statistics of CEMP-no stars (e.g. \citealt{bonifacio2015,yoon2018}). The carbon abundances inferred from observations also depend on the corrections for stellar evolution effects, leading to uncertainties of up to $\sim 0.2\ (0.1)$ in the differential (cumulative) fraction of CEMP-no stars (see e.g. \citealt{placco2014carbon,aguado2017,debennassuti2017,yoon2018}). Nevertheless, such uncertainties do not affect the general trend that we are aiming to reproduce.} 

\begin{figure}
\centering
\includegraphics[width=1\columnwidth]{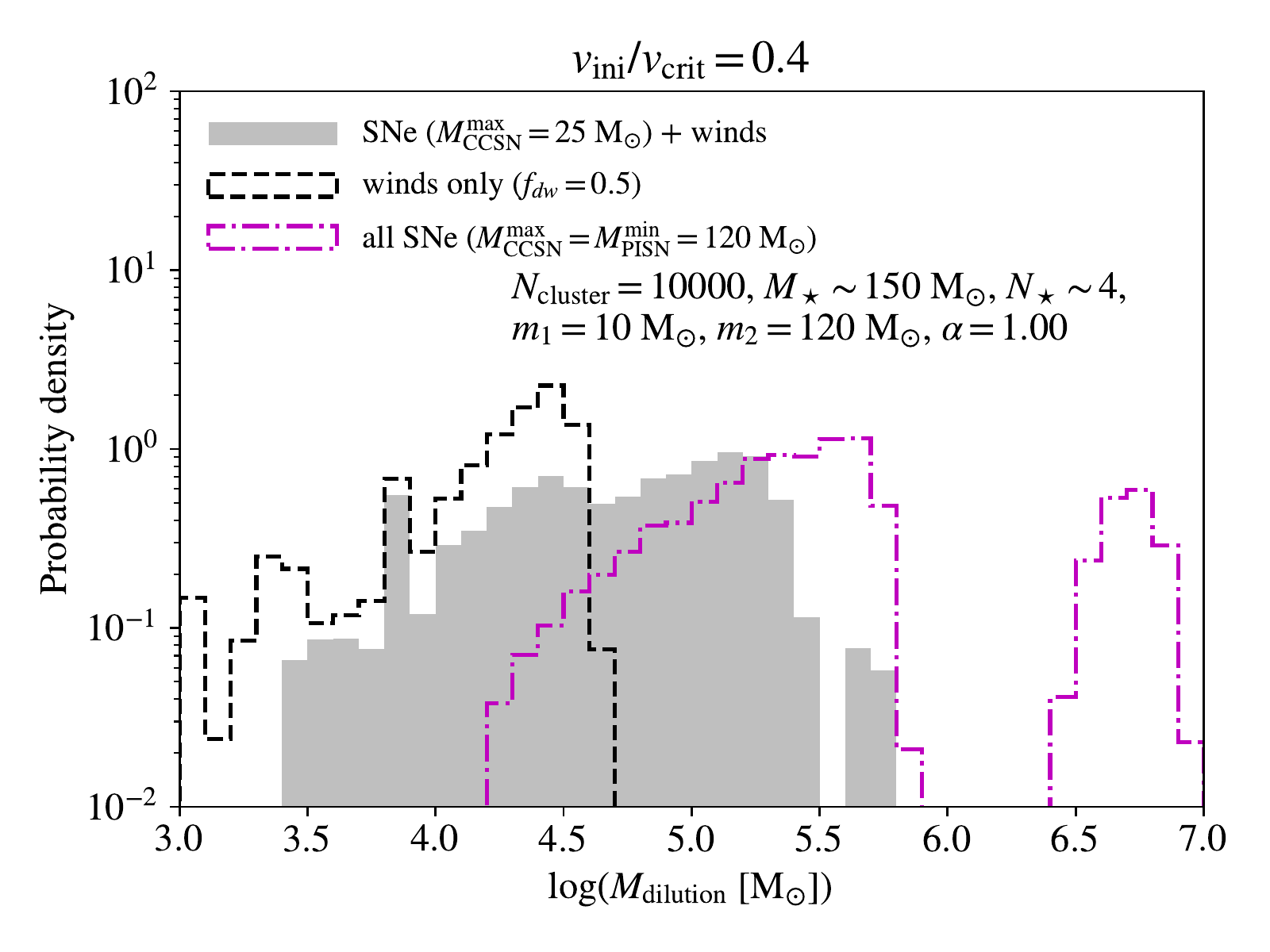}
\vspace{-20pt}
\caption{Dilution mass distributions in self-enriched minihaloes for 3 cases of feedback sources with fiducial SNe + winds (histograms), winds only (dashed contour) and all SNe (magenta contour) under the fiducial IMF and SFE with $f_{dw}=0.5$, each realized with $10^{4}$ clusters/haloes. The stellar evolution grid with rotation $v_{\rm ini}/v_{\rm crit}=0.4$ is adopted.}
\label{r1}
\end{figure}

\begin{figure}
\centering
\includegraphics[width=1\columnwidth]{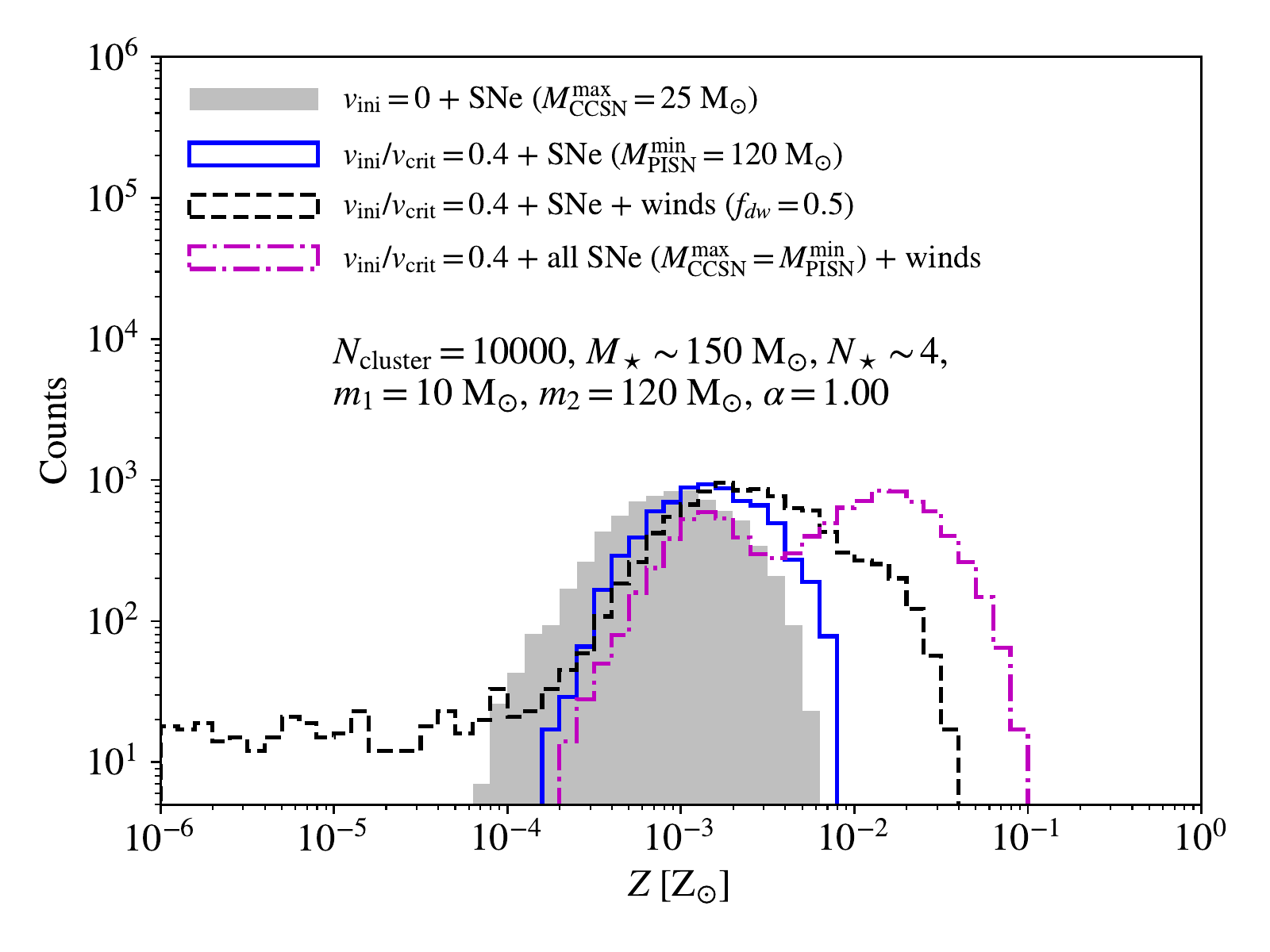}
\vspace{-20pt}
\caption{Metallicity distributions in self-enriched minihaloes for 4 cases of rotation, SN mass range and wind conditions under the fiducial IMF and SFE with $f_{dw}=0.5$, each realized with $10^{4}$ clusters/haloes. The results for all SNe from rotating and non-rotating stellar evolution grids, with and without winds are almost identical, such that only one case is shown here.}
\label{r2}
\end{figure}

\begin{figure*}
    \centering
    \includegraphics[width=1.7\columnwidth]{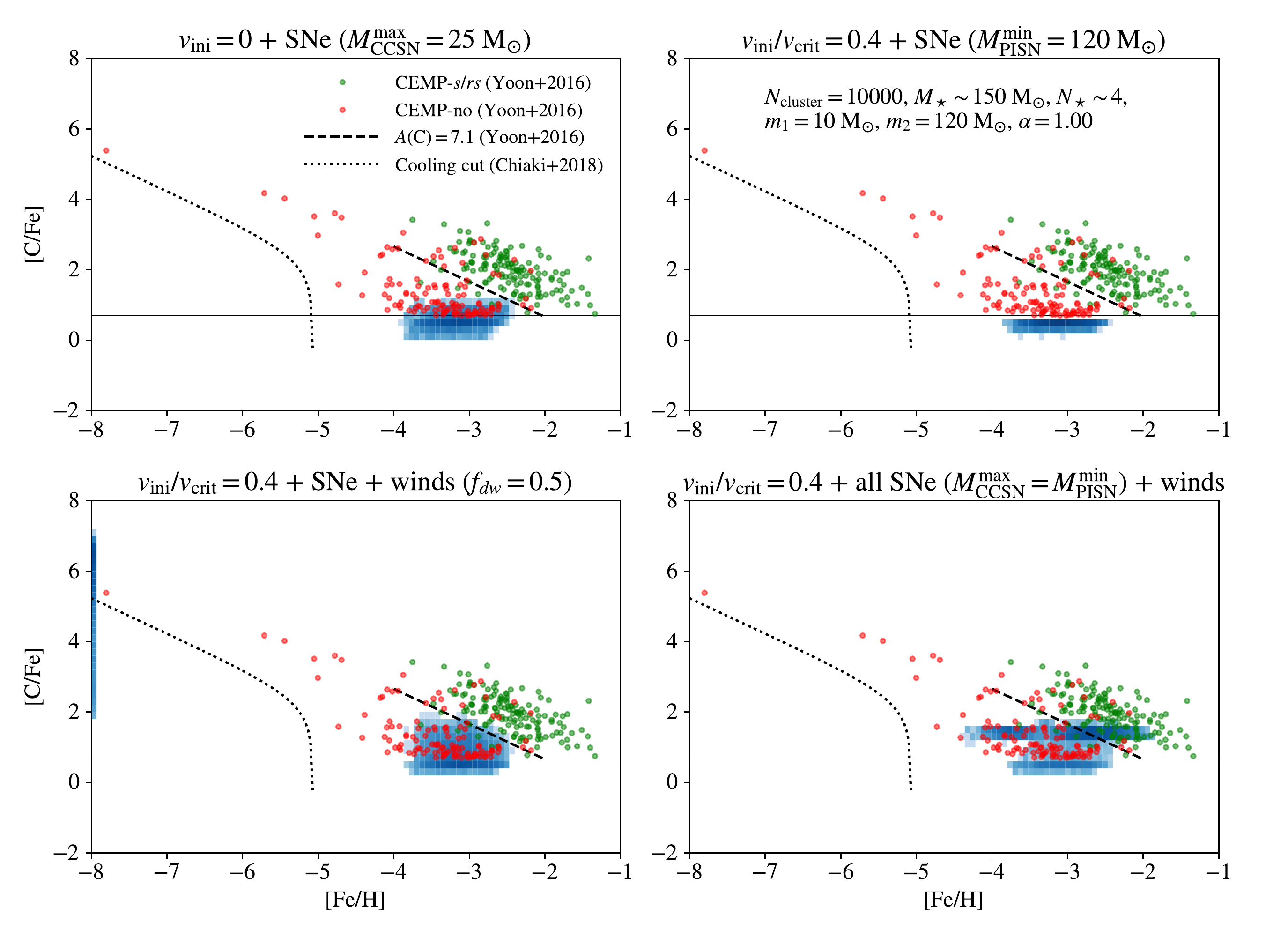}
    \vspace{-10pt}
    \caption{{ Distributions of self-enriched minihaloes (blue pixels) in the [C/Fe]-[Fe/H] space for 4 cases of rotation, SN mass range and wind conditions each realized with $10^{4}$ clusters/haloes under the fiducial IMF and SFE with $f_{dw}=0.5$. We start with the reference non-rotating model with $M_{\rm CCSN}^{\max}=25\ \rm M_{\odot}$, and add rotation, winds and additional SNe (for stars with $m_{\star}>25\ \rm M_{\odot}$) successively from the top-left to bottom-right.} Iron-free samples are shown at the left end with $\rm [Fe/H]=-8$. Observed CEMP-no and CEMP-$s/rs$ stars from \citet{yoon2016observational} are plotted as red and green dots for comparison. We also show the cut $A(\mathrm{C})=7.1$ between CEMP-no and CEMP-$s/rs$ stars proposed by \citet{yoon2016observational} and the (dust) cooling criterion $10^{[\rm C/H]-2.30}+10^{[\rm Fe/H]}>10^{-5.07}$ from \citet{chiaki2017classification} with the dashed and dotted lines, respectively.}
    \label{r3}
\end{figure*}

\subsection{Self-enriched minihaloes}
\label{s3.1}
In this subsection, we consider self-enriched minihaloes with $M_{\rm h}=1.25\times 10^{6}\ \rm M_{\odot}$ at $z=15$ {($T_{\rm vir}\simeq 2000\ \rm K$), as typical birth places of Pop~III stars in the $\Lambda$CDM cosmology. As mentioned in Sec.~\ref{s2.2.4}, our results are actually insensitive to the minihalo mass and redshift such that we restrict our analysis to this specific halo configuration and defer a more detailed investigation to cosmologically representative populations of self-enriched minihaloes in future work.}

\subsubsection{Fiducial IMF and SFE}
\label{s3.1.1}

We start with the situations of fiducial IMF and SFE with $\alpha=1$ and Approach I IMF sampling for $N_{\max}=6$, leading to typical total stellar mass and number of stars as $M_{\star}\sim 150\ \rm M_{\odot}$ and $N_{\star}\sim 4$. We also fixed the wind depth parameter to the fiducial value $f_{dw}=0.5$. Before going into detailed chemical signatures, we would like to highlight some general features of SN and wind feedback (see Sec.\ref{s2.2}) with the distributions of dilution mass for fiducial SNe + winds, winds only and all SNe (including $m_{\star}>25\ \rm M_{\odot}$) models, shown in Fig.~\ref{r1}. Without SNe, we have $M_{\rm dilution}\sim 10^{3}-5\times 10^{4}\ \rm M_{\odot}$ from purely winds. When all stars explode as SNe, we have $M_{\rm dilution}\gtrsim 2\times 10^{4}\ \rm M_{\odot}$, and the effect of winds is almost completely covered up. This shows that SN feedback is stronger than wind feedback in most cases, such that the dilution mass is mostly determined by SN properties as long as SNe are present. Nevertheless, if stars with $m_{\star}>25\ \rm M_{\odot}$ become BHs without SNe, a fraction of clusters will have no SN and be dominated by winds. Under the fiducial IMF and SFE, the mixing regions for the combination of SN and wind feedback have a broad mass range of $\sim 3\times 10^{3}-3\times 10^{5}\ \rm M_{\odot}$, about one quarter of which are dominated by winds. It will be shown below that such systems are vital for reproducing the observed CEMP-no stars. We also find that the dilution mass distribution of the all SNe model has a second peak at a few $10^{6}\ \rm M_{\odot}$, which denote the cases with strong enough SN feedback that can break into the IGM\footnote{This is also expected to be the case when at least one PISN occurs with $E_{\rm SN}\gtrsim 10^{52}\ \rm erg$.}, leading to large SN bubbles, as shown in Fig.~\ref{f3}.

Under the fiducial IMF and SFE, we first look into the effects of rotation, mass range of SNe progenitors and stellar winds on abundance patterns, turning off faint SNe and ISM accretion. As we are mostly concerned with winds triggered by rotation, below we only consider wind enrichment for the rotating models with $v_{\rm ini}/v_{\rm crit}=0.4$. { Actually, the results for winds from the non-rotating models are similar to those shown here for the rotating case. The difference is that there are generally less scatters in carbon enhancement and higher fractions of CEMP stars in the non-rotating case.} 
{Starting from the reference non-rotating model with $M_{\rm CCSN}^{\max}=25\ \rm M_{\odot}$, we add rotation, winds and additional SNe (for stars with $m_{\star}>25\ \rm M_{\odot}$) successively, exploring four cases each with $N_{\rm cluster}=10^{4}$ realizations. The resulting metallicity distributions are shown in Fig.~\ref{r2}. We also consider the situations for all SNe without winds and/or rotation, whose results are similar to the case of rotation + all SNe + winds and thus not shown. } We find that by including winds or additional SNe for $m_{\star}>25\ \rm M_{\odot}$, systems with $Z\sim 0.01-0.1\ \rm Z_{\odot}$ can be produced, which are absent otherwise. For all situations considered here, the majority of haloes have $Z>10^{-4}\ \rm Z_{\odot}$ after the enrichment, such that the second generation of stars formed within them will include low-mass stars ($m_{\star}\lesssim 0.8\ \rm M_{\odot}$) able to survive into the local Universe and be observed as metal-poor stars. 

Interestingly, when winds are combined with fiducial SNe, there is a tiny fraction ($\sim 1$\%) of systems with $Z<10^{-4}\ \rm Z_{\odot}$. The reason is that our rotating stellar evolution model for $m_{\star}=40\ \rm M_{\odot}$ has significant (up to a factor of $10^{-5}$) lower metal yields than other models with $m_{\star}>25\ \rm M_{\odot}$ for $f_{dw}=0.5$ (see Table~\ref{tab:yields}), such that if a cluster only contain stars around $40\ \rm M_{\odot}$, the resulting metallicity is significantly lower. Although these systems are rare and have little impact on our results, they show the possibility of producing extremely metal-poor (but not primordial) environments from wind enrichment with potentially significant scatters in the metal yields from winds. Actually, it is shown in \citet{murphy2021grids} that rotation has complex effects on stellar structure, likely leading to significant scatters in metal yields.

The results on the carbon enhancement ([C/Fe]) and iron abundance ([Fe/H]) diagram are shown in Fig.~\ref{r3}, in comparison to the observational data for CEMP stars from \citet{yoon2016observational}. It turns out that none of these models can sufficiently reproduce the most carbon-enhanced stars with $\rm [C/Fe]\gtrsim 2$ seen in observation. However, except for the rotating model without winds and additional SNe at $m_{\star}>25\ \rm M_{\odot}$ (top-right panel), the majority of observed CEMP-no stars at $\rm -4\lesssim [Fe/H]\lesssim -2$, $\rm [C/Fe]\lesssim 2$ is covered by our predictions. This outcome indicates that normal SNe always produce sufficient iron, such that the most iron-poor and carbon-enhanced stars in observations cannot form from enrichment of normal SNe, which is exactly the motivation of faint SN models. The inclusion of stellar winds alone cannot solve this problem as in most cases winds only enhance the carbon abundance by a factor of a few (depending on the IMF and SN mass range), leaving the iron abundance unchanged. Most systems from our models are below the critical absolute carbon abundance\footnote{The absolute carbon abundance is defined as $A(\mathrm{C})\equiv \log_{10}(n_{\rm C}/n_{\rm H})+12={\rm [C/Fe]+[Fe/H]}+A(\mathrm{C})_{\odot}\simeq {\rm [C/Fe]+[Fe/H]}+8.44$} $A(\mathrm{C})=7.1$ introduced by \citet{yoon2016observational} to distinguish CEMP-no and CEMP-$s/rs$ stars, consistent with the picture that CEMP-no stars trace primordial enrichment while CEMP-$s/rs$ stars are mostly explained by external enrichment in binaries (e.g. \citealt{komiya2020faint}). The difference between rotating and non-rotating models are insignificant when all stars explode as SNe, while if only stars with $m_{\star}<25\ \rm M_{\odot}$ become SNe, the rotation model produces less carbon compared with the rotation model. We also find that winds hardly make any difference when all stars become SNe, which is a natural consequence from the fact that SN feedback is stronger than that of winds (see Fig.~\ref{r1}). 

\begin{figure}
    \centering
    \includegraphics[width=1\linewidth]{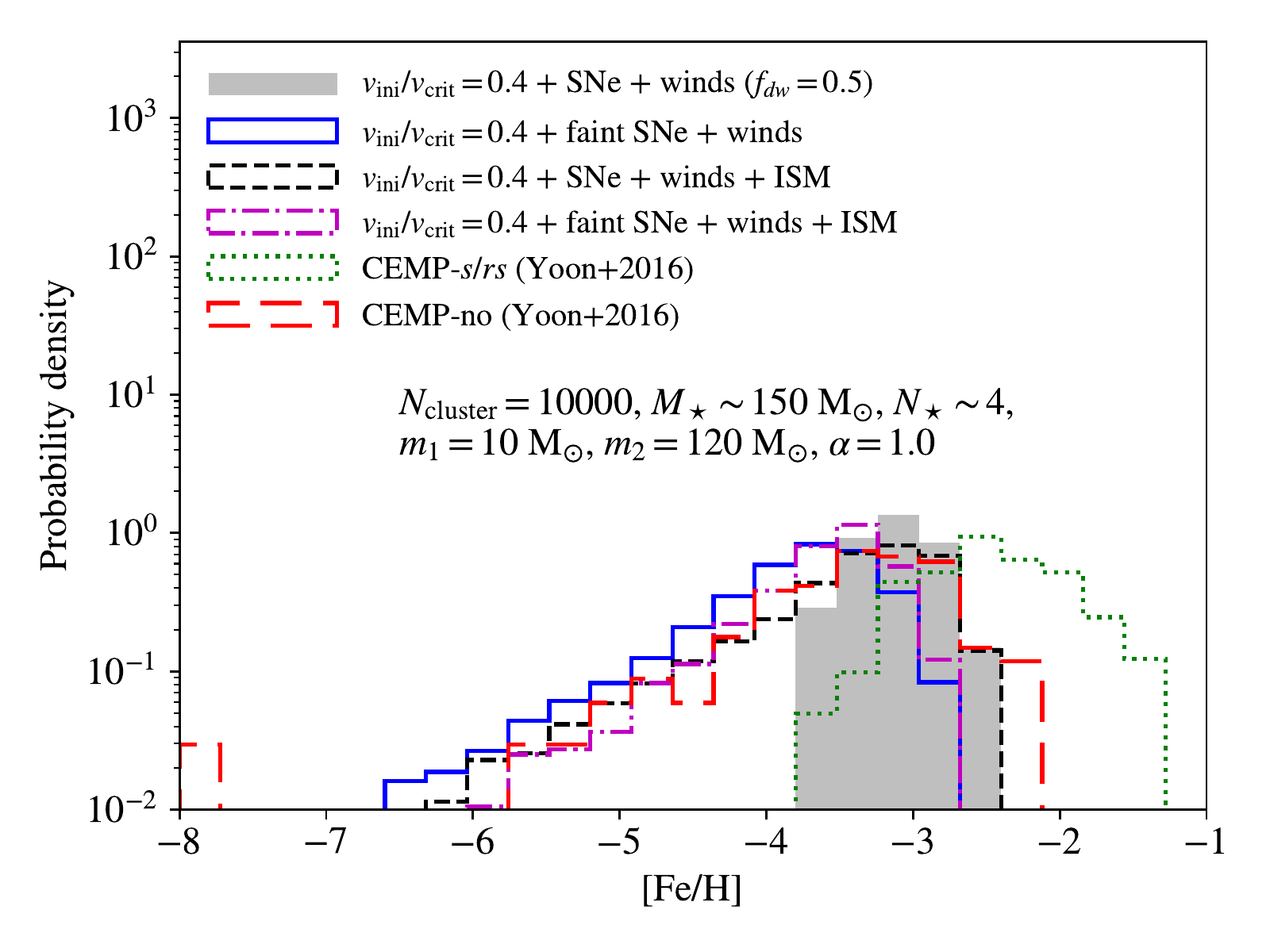}
    \vspace{-20pt}
    \caption{Iron abundance distributions in self-enriched minihaloes (with $\rm [C/Fe]>0.7$) for rotation + winds models under the fiducial IMF and SFE (given $f_{dw}=0.5$), with different treatments for faint SNe and ISM accretion. The long dashed and dotted contours show the observational data for CEMP-no and CEMP-$s/rs$ stars from \citet{yoon2016observational}.}
    \label{r4}
\end{figure}

\begin{figure}
    \centering
    \includegraphics[width=1\linewidth]{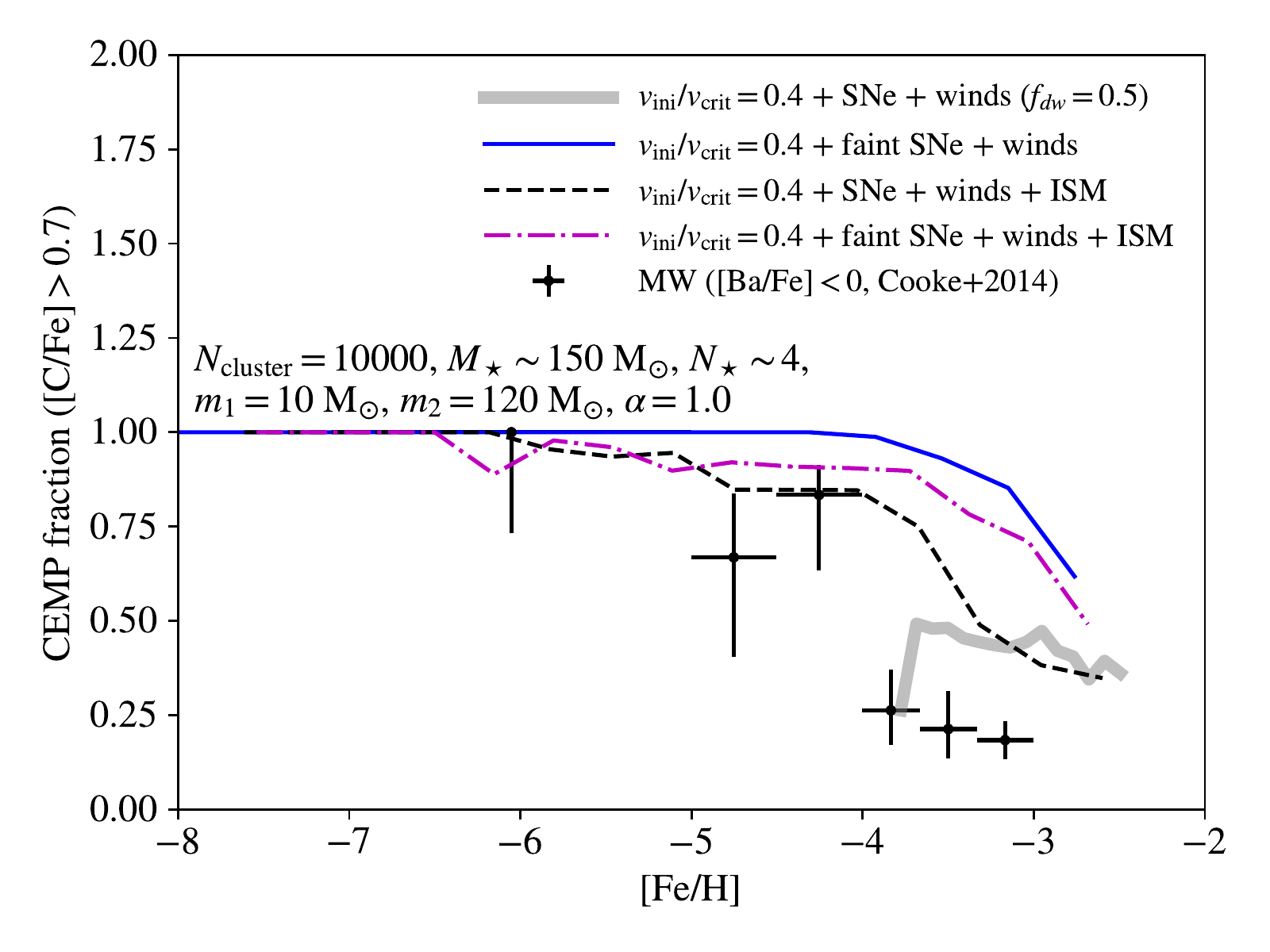}
    \vspace{-20pt}
    \caption{Fractions of CEMP stars as functions of [Fe/H] from self-enriched minihaloes for rotation + winds models under the fiducial IMF and SFE (given $f_{dw}=0.5$), with different treatments for faint SNe and ISM accretion. The observational results for CEMP-no stars ($[\rm Ba/Fe]<0$) in the MW compiled by \citet{cooke2014carbon} are shown with errorbars for comparison.}
    \label{r6}
\end{figure}

\begin{figure*}
    \centering
    \includegraphics[width=1.7\columnwidth]{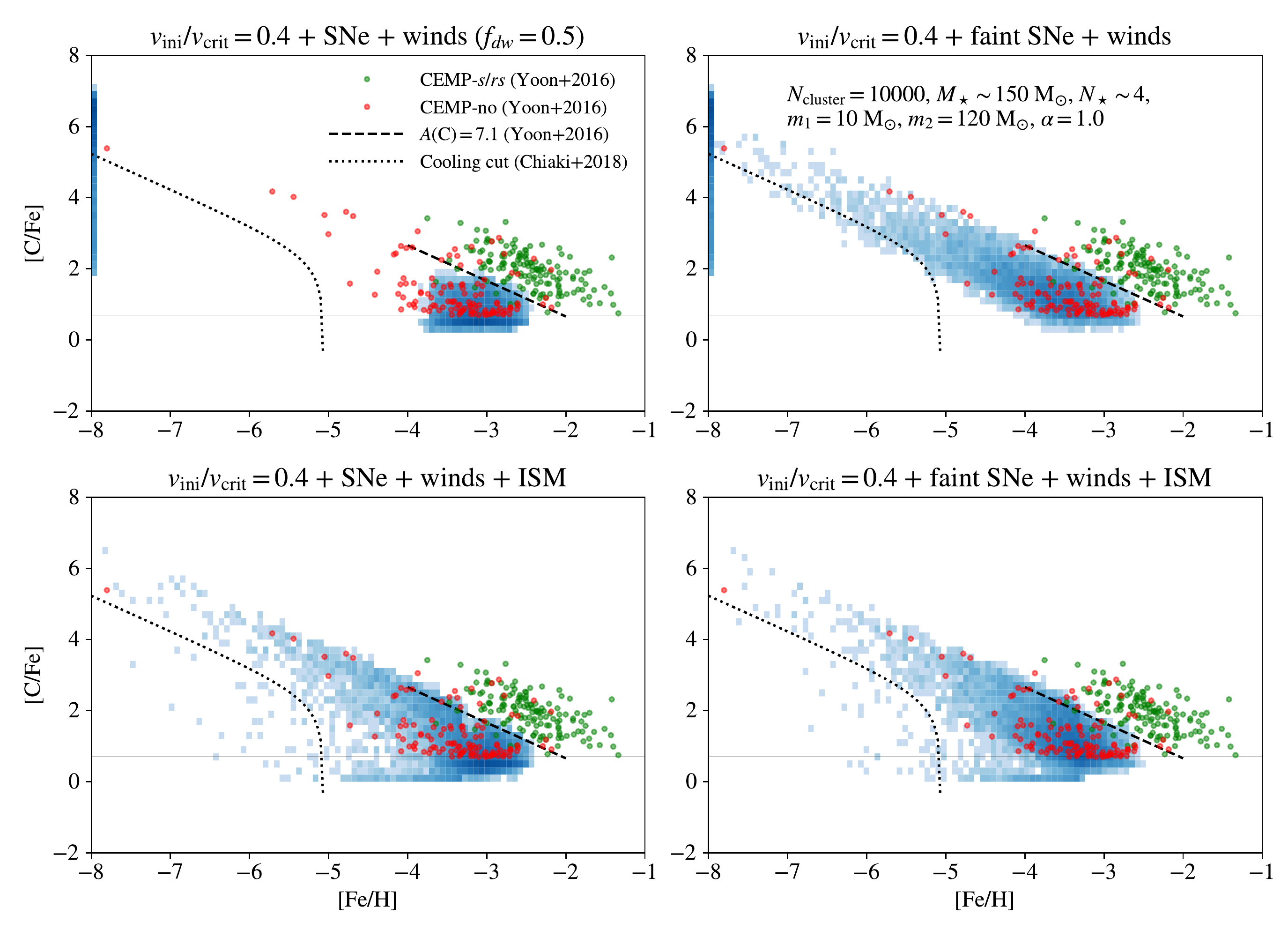}
    \vspace{-10pt}
    \caption{The same as Fig.~\ref{r3} but for 4 cases with and without faint SNe (left to  right) and ISM accretion (top to bottom) under the fiducial IMF and SFE with $f_{dw}=0.5$. }
    \label{r7}
\end{figure*}

As shown in Fig.~\ref{r3}, the key failure of the above models is that the most iron-poor and carbon-enhanced systems are missing. Faint SNe have long been proposed to solve this problem with reduction of iron yields by fallback-mixing {(see e.g. \citealt{umeda2005variations,iwamoto2005,cooke2014carbon,salvadori2015,debennassuti2017,hartwig2019fingerprint,jeon2021role})}. Besides, we notice that when winds are combined with fiducial SNe (from $m_{\star}<25\ \rm M_{\odot}$), about one-quarter of systems dominated by wind enrichment are actually iron-free. If the second generation of stars born in these systems gain trace amount of iron from ISM accretion, they may become extremely iron-poor and carbon-enhanced stars. 

In light of this, we then explore the effects of faint SNe and ISM accretion under the fiducial IMF and SFE. The statistics of carbon and iron abundances for 4 cases (again sampled by $10^{4}$ clusters/haloes) with and without faint SNe and ISM accretion are shown in Figs.~\ref{r4}-\ref{r7}, considering winds with $f_{dw}=0.5$ for the rotating stellar evolution grid. It turns out when \textit{either} faint SNe or ISM accretion is considered, or combining the two effects, the predicted statistics of carbon and iron abundances are generally consistent with observations for CEMP-no stars \citep{cooke2014carbon,yoon2016observational} at $[\rm Fe/H]\lesssim -3.5$ and $A(\mathrm{C})\lesssim 7$. There is a small fraction of CEMP-no stars out of this regime in observations not covered by our models. We expect these relatively iron/carbon-rich stars to be dominated by external enrichment in binaries (see e.g. \citealt{komiya2020faint}). { Actually, a relatively large binary fraction, $47_{-14}^{+15}\%$, is found for carbon-rich ($A(\mathrm{C})\gtrsim 7$) CEMP-no stars in observations \citep{arentsen2019binarity}, indicating that mass transfer in binaries likely plays an important role for producing such chemical signatures. This actually raises an interesting question whether the cut at $A(\mathrm{C})\sim 7$ proposed by \citet{yoon2016observational} divides two groups with different amounts of $s$-elements (CEMP-no vs. CEMP-$s/rs$), or formed in different enrichment scenarios (primordial enrichment vs. binary mass transfer).}

As mentioned in Sec.~\ref{s2.3}, the faint SN and ISM accretion parameters are chosen to match the shape of the observed distribution of [Fe/H] in CEMP-no stars at $[\rm Fe/H]\lesssim -4$. The models with ISM accretion achieve excellent agreements with observations at $[\rm Fe/H]\lesssim -3$ as shown in Fig.~\ref{r4}. {The model with winds + faint SNe only\footnote{We also explore the faint SN models without winds, finding that it is only when all stars explode as (faint) SNe that the most carbon-rich ($A(\mathrm{C})\sim 7$) CEMP-no stars can be reproduced, consistent with the results in \citet{cooke2014carbon}.} tends to slightly overproduce the most iron-poor systems ($[\rm Fe/H]\lesssim -4$) by about 0.2 dex, which is actually close to the observational errors}. In all cases, particularly for faint SNe, stars with $[\rm Fe/H]\gtrsim -3$ are underestimated, implying that the binary channel of forming CEMP-no stars can be important in this regime. 
The trend that the fraction of CEMP-no stars increases with decreasing [Fe/H] is also reproduced (Fig.~\ref{r6}). However, the fraction of CEMP-no stars is over-predicted at $\rm [Fe/H]\gtrsim -4$ in all cases, particularly for the model with winds + faint SNe only. One explanation is that metal-poor stars can also be enriched by Pop~II stars, especially for those with higher [Fe/H], such that they are less likely to be carbon-enhanced. {Actually, semi-analytical models in \citet{salvadori2015,debennassuti2017} and cosmological simulations by \citet{jeon2021role} found that metal-poor stars in this regime ($\rm [Fe/H]\gtrsim -4$) can also form out of Pop~II metals in ultra-dwarf galaxies with and without carbon enhancement, although not as strong as enriched by Pop~III stars.} It will be shown below that considering the contribution of Pop~II stars can indeed achieve better agreements with observation. We also find that almost all haloes in our models satisfy the criterion $10^{[\rm C/H]-2.30}+10^{[\rm Fe/H]}>10^{-5.07}$ based on the observed CEMP stars and the cooling from dust grains \citep{chiaki2017classification}\footnote{We also verify that most of our systems satisfy the cooling criterion for \CII\ and \OI\ lines in \citet{frebel2007probing}.}, i.e. they can indeed host low-mass second-generation stars that will survive to the present. 

Based on the above analysis, we propose a novel channel of forming CEMP-no stars with carbon produced by winds from fast-rotating Pop~III stars and iron coming from ISM accretion, without the need of faint SNe. The predicted carbon and iron abundances of second-generation stars in self-enriched minihaloes from this (rotation +) winds + ISM accretion channel are consistent with those of observed CEMP-no stars (at $[\rm Fe/H]\lesssim -3$ and $A(\mathrm{C})\lesssim 7$) under proper assumptions for ISM accretion. 
In this scenario, carbon-rich but iron-free second-generation stars can form in systems dominated by wind enrichment (without SNe) and gain trace amount of iron from ISM accretion to become the most iron-poor carbon-enhanced stars seen observations with $\rm [Fe/H]\lesssim -4$ and $[\rm C/Fe]\gtrsim 2$.

\begin{figure}
    \centering
    \includegraphics[width=1\linewidth]{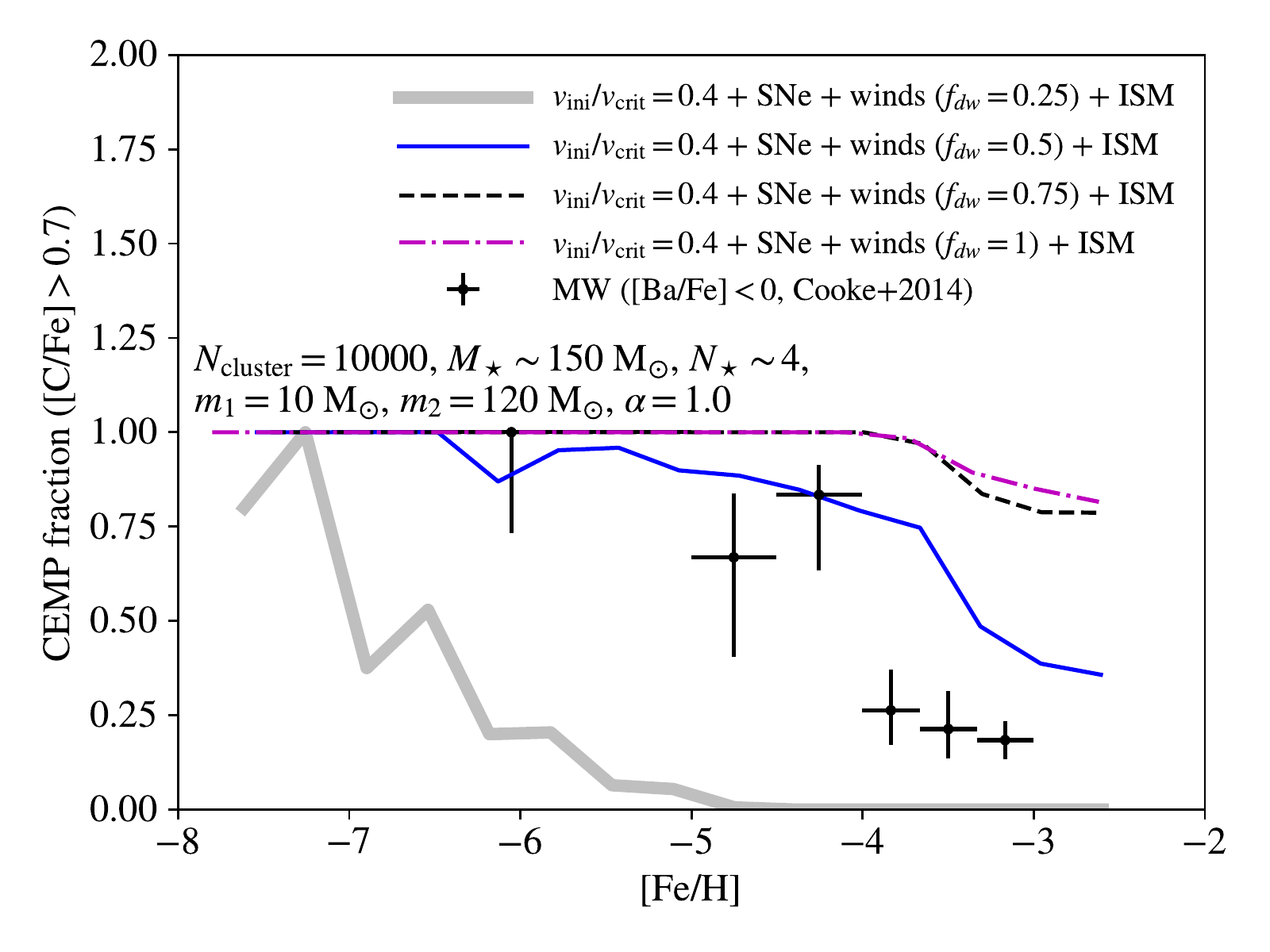}
    \vspace{-20pt}
    \caption{Fractions of CEMP stars as functions of [Fe/H] from self-enriched minihaloes in the the winds + ISM accretion channel with different wind depth parameters: $f_{dw}=0.25$ (thick), 0.5 (solid), 0.75 (dashed) and 1 (dashed-dotted). Again, the observational results for CEMP-no stars ($[\rm Ba/Fe]<0$) in the MW compiled by \citet{cooke2014carbon} are shown with errorbars.}
    \label{r8}
\end{figure}

\begin{figure*}
    \centering
    \includegraphics[width=1.7\columnwidth]{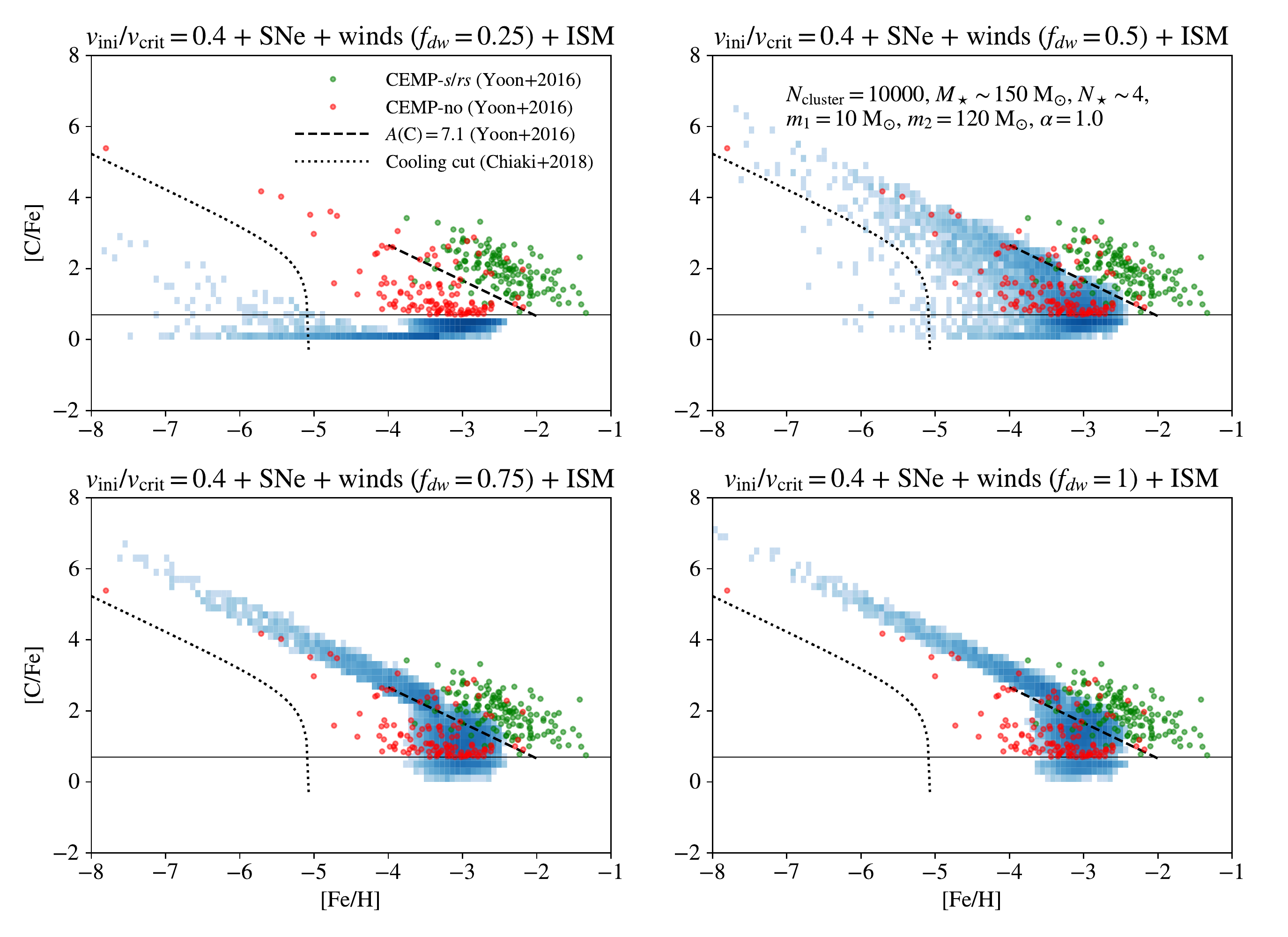}
    \vspace{-10pt}
    \caption{The same as Fig.~\ref{r3} but for the winds + ISM accretion models with different wind depth parameters: $f_{dw}=0.25$ (top-left), 0.5 (fiducial, top-right), 0.75 (bottom-left) and 1 (bottom-right). }
    \label{r9}
\end{figure*}

\newpage

\subsubsection{Parameter Dependence for the winds + ISM accretion channel}
\label{s3.1.2}
{
As shown above, the observed population of CEMP-no stars can be well reproduced with (rotation +) winds + ISM accretion under the fiducial IMF sampling ($N_{\max}=6$, $\alpha=1$, $M_{\star}\sim 150\ \rm M_{\odot}$) and wind depth ($f_{dw}=0.5$). This particular setup of IMF, SFE and $f_{dw}$ is referred as the fiducial model henceforth.} To be concise, below we just focus on the channel of winds + ISM accretion and explore the dependence of abundance patterns on wind depth ($f_{dw}$), IMF ($\alpha$) and SFE ($N_{\max}$ and $M_{\star}$). 

For the wind depth parameter $f_{dw}$, we compare the results for $f_{dw}=0.25$, 0.5, 0.75 and 1 in Fig.~\ref{r8} and \ref{r9}, fixing the IMF and SFE to the fiducial choices. As expected, increasing $f_{dw}$ leads to more carbon yields, such that the fraction of CEMP-no stars is increased across the range of [Fe/H] (Fig.~\ref{r8}), especially for the relatively iron-rich ($\rm [Fe/H]\gtrsim -4$) regime. The [Fe/H] distribution for CEMP stars is also slightly biased towards higher [Fe/H] for $f_{dw}>0.5$, since more iron-rich systems can also become carbon enhanced given more carbon yields. This effect is minor and the distribution of [Fe/H] for CEMP stars is almost identical for $f_{dw}>0.25$ (not shown here). Besides, carbon enhancement [C/Fe] is boosted by higher $f_{dw}$. Actually, for $f_{dw}>0.5$, the carbon-to-iron ratios of systems at $\rm [Fe/H]<-3.5$ predicted by our models are systematically higher than those seen in observations (Fig.~\ref{r9}). We also find that the statistics of carbon and iron abundances change dramatically from $f_{dw}=0.25$ to $f_{dw}=0.5$, showing that significant carbon enhancement from wind enrichment is only activated when winds go deep enough into the star.

\begin{figure}
    \centering
    \includegraphics[width=1\linewidth]{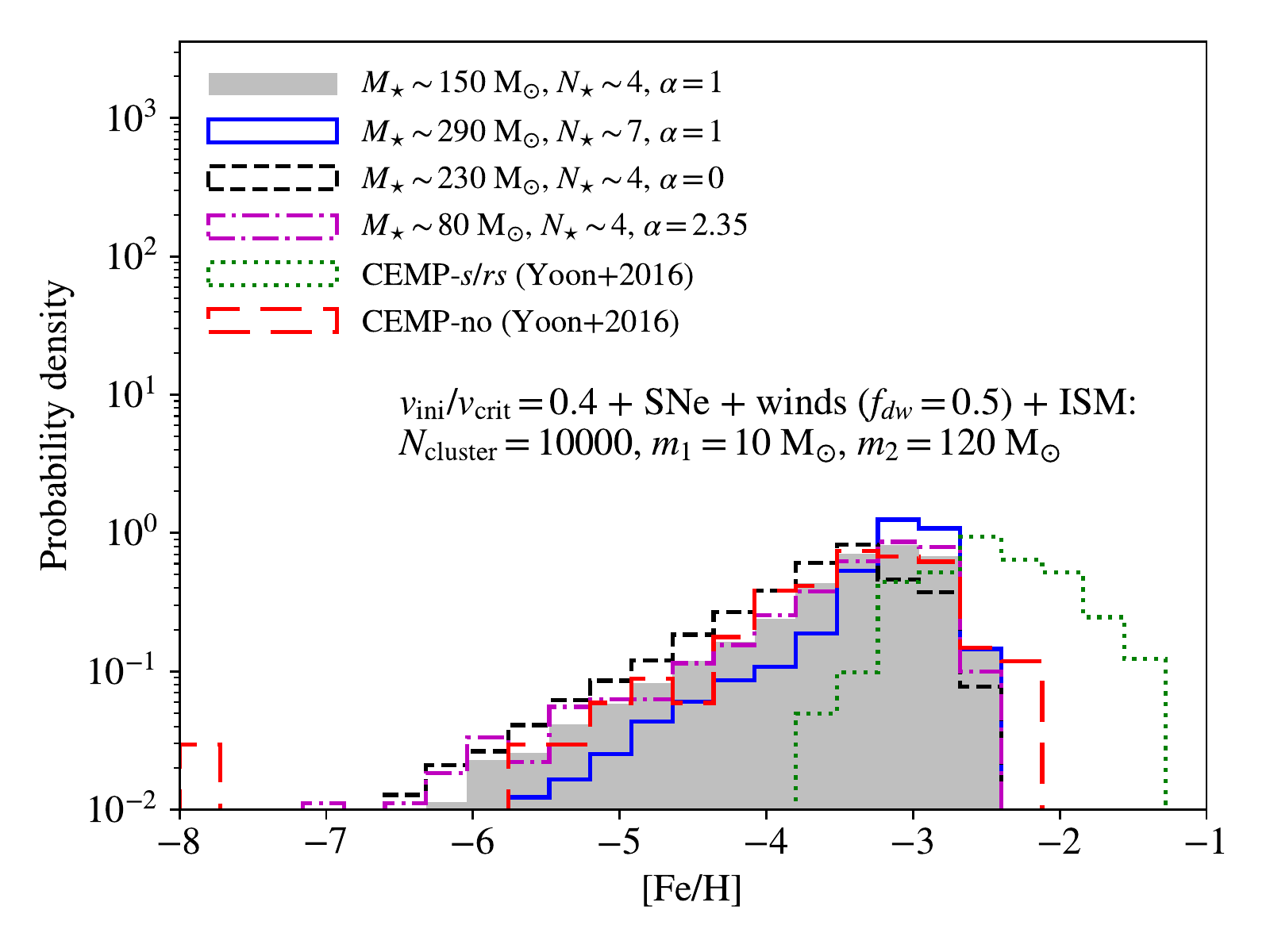}
    \vspace{-20pt}
    \caption{Iron abundance distributions from self-enriched minihaloes (with $\rm [C/Fe]>0.7$) in the the winds + ISM accretion channel for different SFE/star number distributions and IMFs. Models with the fiducial IMF and SFE ($\alpha=1$, $N_{\max}=6$), enhanced SFE/star number ($N_{\max}=12$), top-heavy IMF ($\alpha=0$) and bottom heavy IMF ($\alpha=2.35$) are shown with the gray histograms, solid, dashed and dashed-dotted contours, respectively. The long dashed and dotted contours show the observational data for CEMP-no and CEMP-$s/rs$ stars from \citet{yoon2016observational}.}
    \label{r10}
\end{figure}

\begin{figure}
    \centering
    \includegraphics[width=1\linewidth]{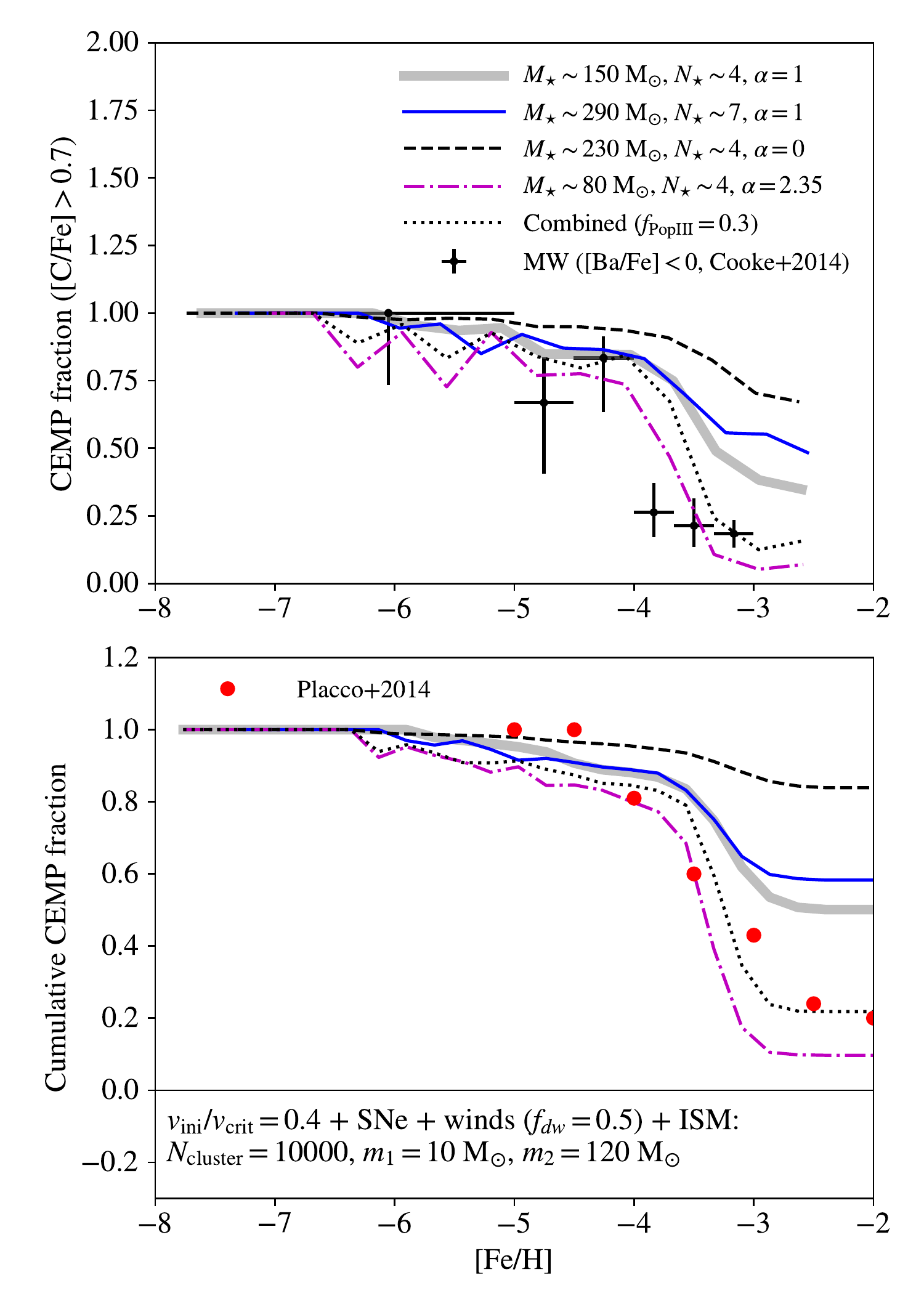}
    \vspace{-20pt}
    \caption{ Differential (top) and cumulative (bottom) fractions of CEMP stars as functions of [Fe/H] from self-enriched minihaloes in the winds+ISM accretion channel for different SFE/star number distributions and IMFs. Models with the fiducial IMF and SFE ($\alpha=1$, $N_{\max}=6$), enhanced SFE/star number ($N_{\max}=12$), top-heavy IMF ($\alpha=0$) and bottom heavy IMF ($\alpha=2.35$) are shown with the thick, solid, dashed and dashed-dotted lines, respectively. The hybrid model combing the fiducial and bottom-heavy cases is shown with the dotted curve, in which realizations of the fiducial model make up 30\% of the samples. On the top panel, the observed differential fractions for CEMP-no stars ($[\rm Ba/Fe]<0$) in the MW compiled by \citet{cooke2014carbon} are shown with errorbars. In the lower panel, we further compare with the cumulative CEMP-no fractions based on high-resolution spectral observations from \citet[see their table~1]{placco2014carbon}, represented by the red data points.}
    \label{r11}
\end{figure}

For IMF and SFE, we again set $f_{dw}=0.5$ and explore 5 new cases in addition to the fiducial model. We first stay within Approach I sampling of the IMF and consider three models, one with increased numbers of stars ($N_{\max}=12$, $\alpha=1$), one with a (more) top-heavy IMF ($\alpha=0$) and one with a (more) bottom-heavy IMF ($\alpha=2.35$) that is similar to the situation of Pop~II and Pop~I stars in the high-mass regime ($m_{\star}>10\ \rm M_{\odot}$). The resulting statistics of carbon and iron abundances in comparison with the fiducial model are shown in Fig.~\ref{r10} and \ref{r11}. For the distribution of [Fe/H] in CEMP stars (Fig.~\ref{r10}), we find that the $\alpha=2.35$ model is almost identical to the fiducial model, while the top-heavy model is biased towards lower [Fe/H]. Increasing star numbers has an opposite effect. 
This trend can be explained by the fact that only relatively low-mass stars with $m_{\star}<25\ \rm M_{\odot}$ will become SNe and provide iron. As mentioned before, since SN feedback is typically stronger than that of winds, one SN in the cluster is enough to reach $\rm [Fe/H]\gtrsim -4$ making the cluster dominated by SN enrichment. Therefore, both higher $\alpha$ and higher $N_{\max}$ can decrease (increase) the chance of having no (at least one) SN in the cluster, and thus decrease (increase) the systems dominated by wind (SN) enrichment. Actually, the fractions of wind-dominated systems are $\sim0.14$, 0.62 and 0.06 for $N_{\max}=12$, $\alpha=0$ and $\alpha=2.35$, respectively, compared with the fiducial value 0.26. For the same reason, the fraction of CEMP-no stars anti-correlates with $\alpha$, but is enhanced by higher $N_{\max}$ at $\rm [Fe/H]\gtrsim -4$ (Fig.~\ref{r11}). The latter trend is caused by the boost of carbon yields in systems with SNe, as the chance of producing a large amount of carbon by winds from massive stars ($m_{\star}>25\ \rm M_{\odot}$) increases with $N_{\rm max}$.

Interestingly, we notice that the fraction of CEMP-no stars is underestimated in the bottom-heavy model with $\alpha=2.35$ at $[\rm Fe/H]>-3.5$. If the metal yields from Pop~II SNe (and winds) are similar to their Pop~III counterparts considered here, our bottom-heavy model can capture the features of Pop~II enrichment (although the detailed normalization of total stellar mass and cosmological context can be different). We then combine the fiducial model with the bottom-heavy model with a free parameter $f_{\rm PopIII}$, which is the fraction of systems from the fiducial model, i.e. enriched by Pop~III stars. {We find that the observed (cumulative) fraction of CEMP-no stars as a function of [Fe/H] can be reproduced within discrepancies $\lesssim 0.2$ (0.1) across the range of $-8\lesssim [\rm Fe/H]\lesssim -3$, given $f_{\rm PopIII}=0.3$. The agreement with observations can be further improved considering that $f_{\rm PopIII}$ likely anti-correlates with [Fe/H] in reality.} This confirms the above speculation that enrichment from Pop~II stars also contributes significantly to the observed population of metal-poor stars, especially at $[\rm Fe/H]\gtrsim -4$. 

{Nevertheless, the transition to the CEMP-dominated regime (with a differential CEMP fraction $\gtrsim 50\%$) in this combined model occurs at $[\rm Fe/H]\simeq -3.6$, higher than in observations ($[\rm Fe/H]\sim -4$), which is also a general feature in our predictions. Actually, this transition is sensitive to the dilution mass distributions. The outcome here implies that the dilution mass is systematically underestimated in our case, which can be solved with lower ionization fluxes, more energetic SNe and more efficient turbulent mixing. The transition in the cumulative CEMP fraction is sharper in our model than in observations. One possible reason is that we only focus on the typical haloes with identical halo masses at a fixed redshift, while in reality metal enrichment happens across a broad range of halo masses and cosmic time. }

\begin{figure}
    \centering
    \includegraphics[width=1\linewidth]{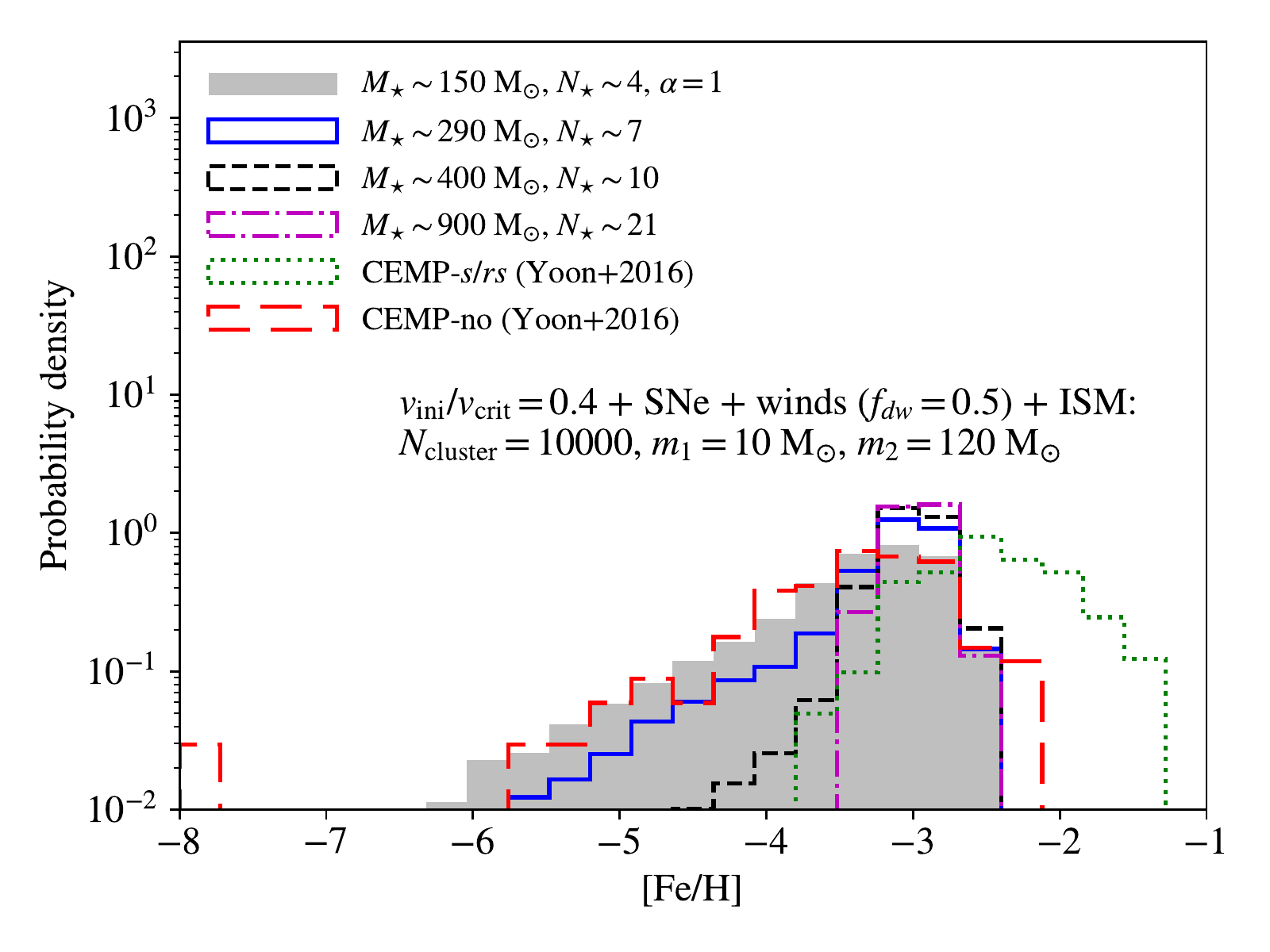}
    \vspace{-20pt}
    \caption{Same as Fig.~\ref{r10} but for different SFEs. Models with $M_{\star}\sim 150$, 290, 400 and $900\ \rm M_{\odot}$ are shown with the gray histograms, solid, dashed and dashed-dotted contours, respectively. }
    \label{r12}
\end{figure}

\begin{figure*}
    \centering
    \includegraphics[width=1.7\columnwidth]{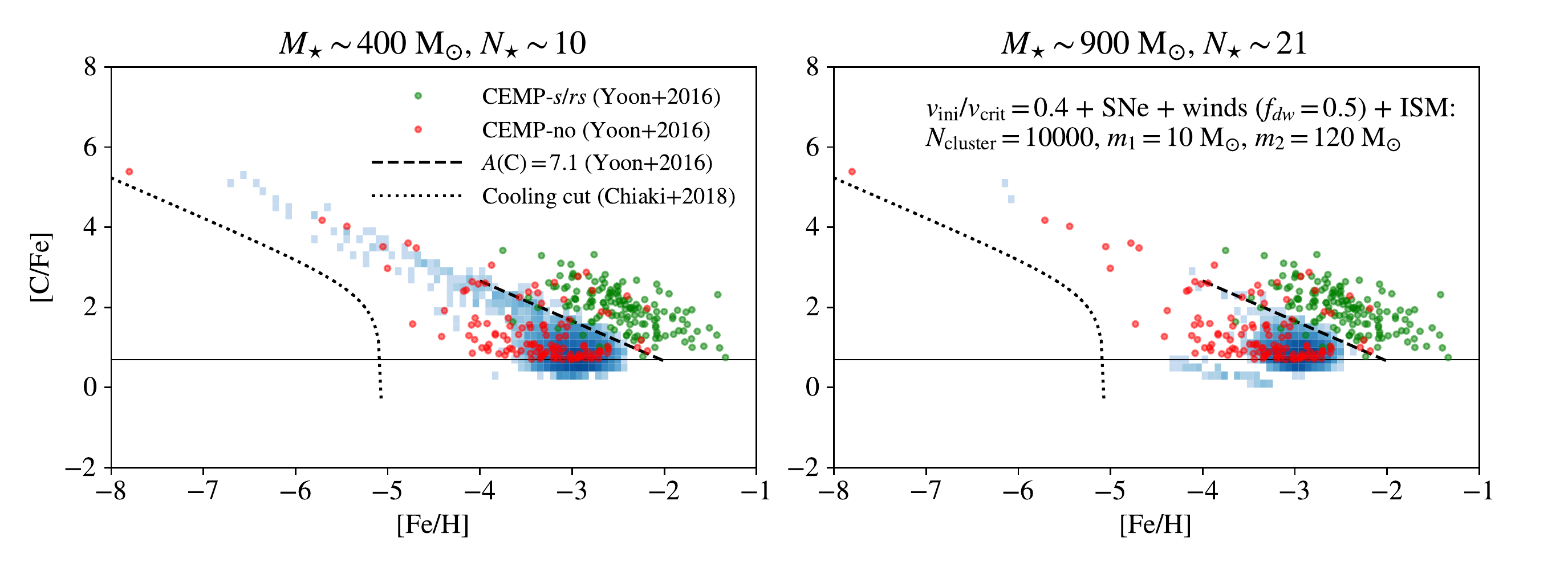}
    \vspace{-10pt}
    \caption{ The same as Fig.~\ref{r3} but for the winds + ISM accretion models from Approach II sampling with different SFEs: $M_{\star}\simeq 400\ \rm M_{\odot}$ (low-SFE, left) and $M_{\star}\simeq 900\ \rm M_{\odot}$ (high-SFE, right), to be compared with the fiducial case with $M_{\star}\sim 150\ \rm M_{\odot}$ in the bottom-left panel of Fig.~\ref{r7}. The distribution in the $M_{\star}\sim 290\ \rm M_{\odot}$ model is similar to that in the fiducial case and thus not shown.}
    \label{r13}
\end{figure*}

Next, we look into the dependence on SFE. In addition to the fiducial model and the model with enhanced star numbers from Approach I IMF sampling, we consider two models from Approach II IMF sampling (still for $\alpha=1$) with pre-determined SFEs: $M_{\star}\simeq 400$ (low-SFE) and $900\ \rm M_{\odot}$ (high-SFE). The results for 4 models with $M_{\star}\sim 150$, 290, 400 and $900\ \rm M_{\odot}$ are shown in Fig.~\ref{r12} and \ref{r13}. Similar to the trend mentioned above (Fig.~\ref{r10}), we find that increasing the SFE will reduce the fraction of extremely iron-poor CEMP stars ($[\rm Fe/H]\lesssim -4$), as shown in Fig.~\ref{r12}. This again results from the correlation between star numbers and fraction of wind-dominated systems, since the number of stars is roughly proportional to the SFE given a fixed IMF. We find that the wind-dominated fraction drops from the fiducial value 0.26 to 0.036 and 0.0011 for the low-SFE and high-SFE models, respectively. The distribution of $[\rm Fe/H]$ for CEMP stars also becomes more concentrated with higher SFE. The reason is simply that higher SFE under a fixed IMF leads to more stars and better sampling of the IMF for each cluster, reducing the scatters among different realizations. This trend is also seen in the [C/Fe]-[Fe/H] space (Fig.~\ref{r13}). 

Actually, the high-SFE model fails to reproduce enough extremely iron-poor CEMP stars (with $\rm [Fe/H]\lesssim -4$) seen in observations, even when faint SNe are considered. The reason is that if Pop~III clusters contain too many members, the chance of having extremely low iron yield from them tends to be low. This implies that if the observed CEMP-no stars at $\rm [Fe/H]\lesssim -4$ are indeed second-generation stars enriched by Pop~III stars, their existence already puts constraint on the SFE of Pop~III star formation\footnote{The detailed constraint on Pop~III SFE depends on the IMF and SN properties. }
We also notice the emergence of a small substructure in Fig.~\ref{r13} around $-4.5\lesssim [\rm Fe/H]\lesssim -3.5$ and $[\rm C/Fe]\lesssim 1$ for the high-SFE model with $M_{\star}\simeq 900\ \rm M_{\odot}$. Similar structures also occur when all stars becoming SNe under the fiducial IMF and SFE (bottom panels of Fig.~\ref{r3}). This subgroup is made of haloes in which the SN feedback is strong enough to break into the IGM with much larger SN bubbles than the confined case, as illustrated in Fig.~\ref{f3} and \ref{r1}., which becomes more important when more Pop~III clusters undergo energetic SN explosions from a higher SFE and/or SN rate.

\begin{figure*}
    \centering
    \includegraphics[width=1.7\columnwidth]{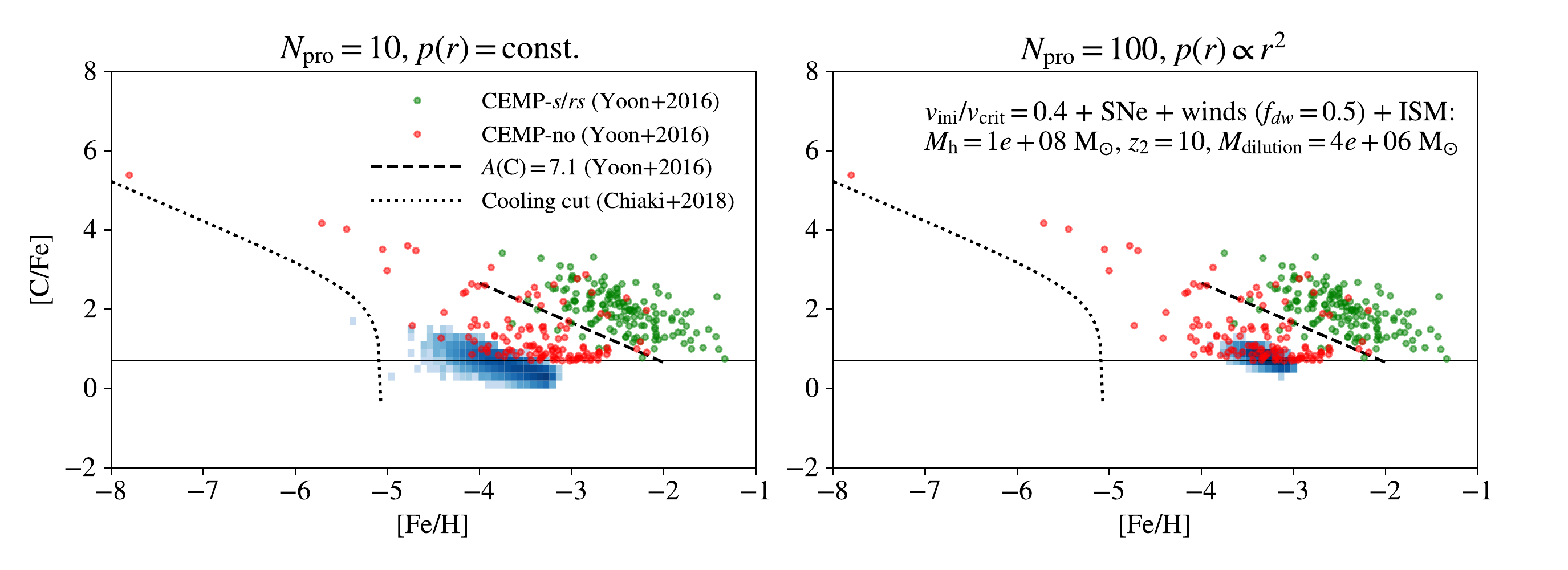}
    \vspace{-10pt}
    \caption{Distributions of central galaxies in AC haloes (blue pixels) in the [C/Fe]-[Fe/H] space for different progenitor numbers and distributions around the halo center. The observational data shown here are explained in Fig.~\ref{r3}.}
    \label{r14}
\end{figure*}

\subsection{Atomic-cooling haloes}
\label{s3.2}

In this subsection, we briefly discuss our results for AC haloes with $M_{\rm h}=10^{8}\ \rm M_{\odot}$ at $z=10$. Note that here we only consider the metal enrichment of the central metal mixing region (i.e. the central galaxy), while satellite haloes may also form second-generation stars before being destroyed. We expect the chemical signatures of such satellite haloes to be similar to those of self-enriched minihaloes. {We again focus on the winds + ISM scenario with the fiducial IMF, SFE and wind depth for the minihaloes that merge into the AC halo, and consider progenitor numbers in the range $N_{\rm pro}\sim 10-100$. We put the minihaloes randomly around the halo center (within the Lagrangian region) following a SIS ($p(r)=\rm const.$) or an uniform distribution ($p(r)\propto r^{2}$). For simplicity, we fixed the halo mass to $1.25\times 10^{6}\ \rm M_{\odot}$ for minihaloes falling into the AC halo. Again, $10^{4}$ haloes are simulated for each case, in which the Pop~III clusters involved for each halo are randomly chosen from $10^{4}$ realizations of the fiducial winds + ISM  model. 

Fig.~\ref{r14} shows two examples for the distributions of AC haloes in the [C/Fe]-[Fe/H] space with $M_{\rm dilution}\simeq 4\times 10^{6}\ \rm M_{\odot}$, i.e. uniform distribution of gas within the mixing region of the central galaxy.} The results for $M_{\rm dilution}\simeq 10^{7}\ \rm M_{\odot}$ (for concentrated gas following a SIS ) are similar and thus not shown. It turns out that AC haloes can only reproduce the observed CEMP-no stars with mild carbon enhancement ($[\rm C/Fe]\lesssim 2$) at $-5\lesssim [\rm Fe/H]\lesssim -3$. Increasing $N_{\rm pro}$ or the concentration of minihaloes that (potentially) contribute to the central mixing region increases the overall metallicity and iron abundance, but cannot provide enough carbon enhancement. Considering faint SNe still cannot solve this problem. This implies that for the central galaxies in more massive haloes where metals are likely from an increasing number of progenitors (and thus individual stars), it is more difficult to increase carbon abundance without increasing iron abundance. In other words, carbon enhancement originates from scatters in the carbon and iron yields from individual metal sources, which can only be preserved when star formation and metal enrichment happen at small scales (e.g. minihaloes), or metal mixing is highly inhomogeneous (see e.g. \citealt{hartwig2019formation}).

\section{Summary and discussions}
\label{s4}
We combine stellar evolution grids for fast-rotating Pop~III stars \citep{murphy2021grids} with a semi-analytical model of stellar feedback and metal enrichment to study the chemical signatures of Pop~III stellar winds. We explore a large parameter space of Pop~III star formation and feedback, including the IMF, SFE, as well as {yields from winds and SNe} for both self-enriched minihaloes ($M_{\rm h}\sim 10^{6}\ \rm M_{\odot}$, $z\sim 15$) and AC haloes ($M_{\rm h}\sim 10^{8}\ \rm M_{\odot}$, $z\sim 10$), the typical hosts of Pop~III stars. We compare our results with the observed population of CEMP-no stars {\citep{spite2013,cooke2014carbon,placco2014carbon,bonifacio2015,yoon2016observational,yoon2018}} as the bona-fide second-generation stars, specifically considering the carbon enhancement and iron abundance space. Here are our main findings for self-enriched minihaloes:
\begin{itemize}
    \item Feedback from stellar winds is usually weaker than that from SNe. That is to say, the masses of metal mixing regions produced only with winds are generally lower compared with the cases when SNe are present. Moreover, a single normal SN is enough to achieve $[\rm Fe/H]\gtrsim -4$ in the enriched medium, rendering the system dominated by SN enrichment.
    \item There is a channel of forming CEMP-no stars with Pop~III stellar winds and surface pollution by accretion from the ISM, without the need for faint SNe, when {massive Pop~III stars ($m_{\star}\gtrsim 25\ \rm M_{\odot}$) collapse into BHs without SNe and} a significant fraction of metals produced in the stellar core can be lost in winds ($f_{dw}\gtrsim 0.5$). In this scenario, carbon-rich but iron-free second-generation stars can form in systems dominated by winds, gaining iron by ISM accretion to become the most iron-poor and carbon-enhanced stars seen in observations ($[\rm Fe/H]\lesssim -4$, $[\rm C/Fe]\gtrsim 2$). The statistics of carbon and iron abundances in the observed population of CEMP-no stars can be well reproduced when the ISM accretion contribution of iron is constrained by the observed $[\rm Fe/H]$ distribution of CEMP-no stars.
    \item {Considering metal enrichment from Pop~II stars (with a more bottom-heavy IMF compared with Pop~III) is necessary to reproduce the observed fraction of CEMP stars as a function of $\rm [Fe/H]$ in the MW \citep{cooke2014carbon,placco2014carbon}, especially in the relatively metal-rich regime with $[\rm Fe/H]\gtrsim -4$. Actually, without a significant ($\sim 70\%$) contribution of Pop~II enrichment, the CEMP fraction will be overestimated by a factor of a few at $[\rm Fe/H]\gtrsim -4$. This is consistent with the results in semi-analytical models \citep{salvadori2015,debennassuti2017} and cosmological simulations \citep{jeon2021role}, which indicate that Pop~II enrichment in massive minihaloes and AC haloes can form C-normal stars efficiently at $\rm [Fe/H]\gtrsim -4.5$. } 
\end{itemize}
For AC haloes, we only consider the central mixing region corresponding to the central galaxy and find that
\begin{itemize}
    \item CEMP-no stars with the most carbon enhancement ($[\rm C/Fe]\gtrsim 2$) cannot form in the central galaxies of AC haloes, regardless of the choices of feedback parameters and number of progenitor minihaloes that contribute to metal enrichment. Increasing the progenitor number results in smaller scatters of carbon and iron abundances.
    \item This implies that carbon enhancement in metal-poor stars originates from scatters in the carbon and iron yields from individual sources of metals. These scatters can only be captured by second-generation stars when metal enrichment (once) occurs in small structures {(see e.g. \citealt{salvadori2015})} or is highly inhomogeneous (see e.g. \citealt{hartwig2019formation}).
\end{itemize}

In general, our analysis for the first time verifies the potentiality for metal enrichment from stellar winds of fast-rotating Pop~III stars to explain the observed CEMP-no stars in the typical sites of Pop~III star formation, {achieving a comparable level of agreement with observations as previously well-explored scenario of faint SNe (e.g. \citealt{cooke2014carbon,salvadori2015,sarmento2016following,debennassuti2017,sharma2018origins,hartwig2019fingerprint,chiaki2020seeding,jeon2021role}). Note that our models do not exclude alternative scenarios for the formation of CEMP-no stars.} In nature, a variety of processes can occur and even work together, such that Pop~III stars can have strong winds, die in faint SNe and also undergo close binary interactions. Considering the large uncertainties in Pop~III properties, it remains an open question how the imprints from different pathways of metal enrichment can be distinguished and their relative importance evaluated.

{
Finally, we would like to point out the caveats in our exploratory work: 
\begin{itemize}
    \item We have adopted simplified assumptions under spherical symmetry for some key physical processes and parameters, such as the wind luminosity and composition, distribution of SN energy, surface pollution from ISM accretion and fallback-mixing of faint SNe. We also ignore the effects of binaries, which can alter the wind properties and SN remnants. Particularly, wind enrichment is important only when metals produced in the inner part of the star ($f_{dw}\gtrsim 0.5$) can be carried out by winds, which can result from very efficient internal mixing \citep{meynet2006early,hirschi2007very,ekstrom2008effects,EkFIRST2008}, and/or significant mass loss due to pulsations, approach to the critical velocity and potential instabilities occurring before the core collapse \citep{smith2006role,van2008numerical, Yoon2010,Fuller2017, Fuller2018,Fuller2020, Fuller2021}. Such processes are not explicitly modelled in our stellar evolution grids, and it remains uncertain whether they can occur in massive Pop~III stars. Achieving a better understanding of these mechanisms will be key to more realistically model Pop~III wind feedback. 
    \item Our winds + ISM accretion channel of forming CEMP-no stars only works when a non-negligible fraction of Pop~III stars directly collapse into BHs. Otherwise, if all massive Pop~III stars ($m_{\star}\gtrsim 10\ \rm M_{\odot}$) explode as SNe, the wind signature would be hidden, such that reduction of the iron yields from SNe by fallback-mixing is necessary to form CEMP-no stars. In our fiducial model, we assume that stars with $m_{\star}>25\ \rm M_{\odot}$ become BHs directly without SN explosions based on \citet[see their equs.~51-53]{tanikawa2020fitting}. However, the existence and exact (initial) mass range for such `direct-collapse' Pop~III stars is still uncertain in stellar evolution models (see e.g. fig.~1 in \citealt{Heger2003} and fig.~6 in \citealt{heger2010nucleosynthesis}). 
    \item We do not take into account the cosmological context to self-consistently combine the enrichment from Pop~III and Pop~II stars in a representative population of haloes. Instead, we only consider the `typical' minihaloes and AC haloes with typical halo properties, and introduce the effect of Pop~II stars (modelled by a bottom heavy IMF) with a free weighting parameter $f_{\rm PopIII}$, denoting the fraction of systems dominated by Pop~III enrichment. In this way, we are able to roughly reproduce the relation between the (cumulative) CEMP fraction and $\rm [Fe/H]$ within uncertainties of $\lesssim 0.2$ (0.1), given the best-fit value $f_{\rm PopIII}=0.3$. 
    \item We only consider the effects on (surface) iron and carbon abundances from ISM accretion with a parametric stochastic model calibrated to the observed [Fe/H] distribution of CEMP-no stars at $\rm [Fe/H]\lesssim -4$, and defer the investigation of the full abundance patterns of CEMP-no stars to future work. In reality, the effects of ISM accretion are rather complex, depending not only on the accretion history of second-generation stars but also on the diffusion process that can have different efficiencies for different elements. In our case, the increase of (surface) metallicity from ISM accretion ($\Delta Z<5\times 10^{-4}\ \rm Z_{\odot}$) is mostly smaller than the initial enrichment ($Z\sim 10^{-3}\ \rm Z_{\odot}$) from Pop III stars. Since accretion occurs over a very long timescale ($\sim 10\ \rm Gyr$), the composition of accreted materials (on average) will be similar to the solar composition. Under such conditions, we can safely conclude that for any feature involving elements that are unique or enhanced in Pop III enrichment with respect to the ISM, enriched mostly by Pop II/I stars, accretion has little impact. One example is the C/O ratio, as both C and O are enhanced. For under-abundant elements from Pop~III enrichment, they will be less under-abundant after ISM accretion. This effect can only be derived from detailed hydrodynamic calculations of the stellar surface layer (e.g. \citealt{henkel2018,deal2021}), beyond the scope of this study. 
\end{itemize}

In light of the above caveats, future studies should use more advanced stellar evolution grids with higher initial rotation speeds, additional mass loss processes, and different SN explosion models for a range of initial metallicities ($Z\lesssim 10^{-4}\ \rm Z_{\odot}$). They may also consider the effects of close binary interactions (e.g. \citealt{kinugawa2014possible,belczynski2017likelihood,tanikawa2020merger}). Furthermore, more realistic modelling of the feedback process has to involve 3D~hydrodynamics (e.g. \citealt{rogers2013feedback,geen2015detailed,fierlinger2016stellar}). Furthermore, one needs to fully take into account the cosmological context of metal enrichment with halo merger trees (e.g. \citealt{salvadori2015,hartwig2019fingerprint,debennassuti2017,komiya2020faint}) or cosmological simulations (e.g. \citealt{sarmento2016following,sharma2018origins,chiaki2020seeding,jeon2021role}), to better model ISM accretion and evaluate the chemical signatures of Pop~III stellar winds, thus enabling more direct comparisons with observations for a broader range of elements.
}

\section*{Acknowledgments}
The authors thank Laura Murphy for having provided access to the numerical tables describing the Pop III stellar models. 
Boyuan Liu would like to thank Duo Xu for helpful discussion on wind bubbles and Anke Arentsen for suggestions on the fraction and binarity of CEMP-no stars in observations. 
Georges Meynet has received funding from the European Research Council (ERC) under the European Union's Horizon 2020 research and innovation program (grant agreement No 833925, project STAREX). 

\section*{Data availability}
The data underlying this article will be shared on reasonable request to the corresponding authors.

\bibliographystyle{mnras}
\bibliography{ref} 


\label{lastpage}
\end{document}